\documentclass[aps,prd,%twocolumn,
eqsecnum,showpacs,nofootinbib]{revtex4}

\usepackage{amsmath,amsfonts,amssymb,color,graphicx,latexsym,theorem,mathrsfs,color,bm,
multirow}

\newcommand{\ma}[1]{\mbox{$\mathcal{#1}$}}

\newcommand{\bs}[1]{\mbox{$\boldsymbol{#1}$}}

\newcommand{\D}{{\rm d}}

\newcommand{\ti}{\tilde}

\usepackage{comment}

\begin{document}
%\thispagestyle{empty}

%%%%%%%%%%%%   Title   %%%%%%%%%%%%%
\title{
All the three dimensional Lorentzian metrics admitting three Killing vectors
}

%%%%%%%%%%%%   Authors   %%%%%%%%%%%%%
\author{Masato Nozawa$^1$ and Kentaro Tomoda$^2$}
%\email{masato.nozawa_at_yukawa.kyoto-u.ac.jp\\k-tomoda_at_sci.osaka-cu.ac.jp}

%%%%%%%%%%%%   Address   %%%%%%%%%%%%%
\address{ 
$^1$Center for Gravitational Physics, Yukawa Institute for Theoretical Physics, Kyoto University, Kyoto 606-8502, Japan\\ 
$^2$Advanced Mathematical Institute, Osaka City University, Osaka 558-8585 Japan}

\date{\today}

%%%%%%%%%%%%%%%%%%%%%%%%%%%%%%%%%%%%%%
%                                                                                                %
%                                      Abstract                                          %
%                                                                                                %
%%%%%%%%%%%%%%%%%%%%%%%%%%%%%%%%%%%%%%%
\begin{abstract} 
We obtain all the three-dimensional Lorentzian metrics which admit three Killing vectors. The classification has been done with the aid of the formalism which exploits the obstruction criteria for the Killing equations recently developed by present authors. The current classification method does not rely on the transitivity property of the isometry group.  It turns out that the Lorentzian manifold harbors a much richer spectrum of metrics with various Segre types, compared to the Riemannian case. 
\end{abstract}

%\pacs{} 

\maketitle

%%%%%%%%%%%%%%%%%%%%%%%%%%%%%%%%%%%%%%
%                                                                                                %
%                                   Introduction                                         %
%                                                                                                %
%%%%%%%%%%%%%%%%%%%%%%%%%%%%%%%%%%%%%%

\section{Introduction and summary}

%% given (M, g) -> # of KVs with using diff. invs.
A Killing vector (KV) on a manifold $M$ with a (pseudo-)Riemannian metric $g$ describes a vector field whose flow preserves the metric. 
The isometry group generated by the KVs is of fundamental touchstone in Lorentzian signature, since it has a direct link to
the construction of exact solutions to Einstein's equations.  The importance of KVs in general relativity is even more highlighted by the relevance to 
the globally conserved quantities and constants of geodesic motion. 

Despite the above-mentioned importance of KVs, the existence of any KV for a given a (pseudo-)Riemannian manifold $(M, g)$ is not instantly evident.
A patient difficulty for this issue is how to find, if any, the explicit form and exhaustive list of KVs. 
More than a century ago, Darboux has put forward an inventive idea to determine the exact number of linearly independent KVs 
in two dimensions \cite{Darboux:1894}.  His criterion for the local existence of KVs has been formulated in terms of 
differential invariants and has provided a direct interplay between local geometric invariants and solvability of an overdetermined PDEs, namely the Killing equations. Recently we have witnessed considerable developments in this direction \cite{Kruglikov:2008,Bryant:2008}. 
Notably, the complete set of local obstructions to the existence of KVs in three dimensions has been identified 
in a coordinate invariant fashion both for Riemannian \cite{Kruglikov:2018qcn} and for Lorentzian \cite{Nozawa:2019dwu} signatures, 
which augmented a revival interest for Darboux's methodology. 
The obstructions can be applied to count the exact number of KVs, resulting in severe restrictions on the curvature of $M$. 
This method has overcome some shortcomings inherent in other prescriptions proposed in the literature \cite{Cartan:1946,KarlhedeMaccallum:1982}
in that classification is indeed tractable. 

%% # of KVs with diff. invs. -> a class of (M, g) : 6, 4 KVs review 1
Conversely, one might ask whether one can determine the local form of $g$ by prescribing the obstructions and/or the curvature of $M$. 
Indeed, the existence of KVs has traditionally been used to classify the manifolds, see e.g. \cite{Stephani:2003tm} for a review. 
A three-dimensional (pseudo-)Riemannian manifold is said to be maximally symmetric if it has six KVs, or equivalently, if it has constant sectional curvature. Ricci \cite{Ricci:1898} and Bianchi \cite{Bianchi:1898} have proved that the next maximum number of KVs is four, and have 
carried out the complete classification of all possible local forms of $g$ with four KVs. 
The pursuit for the Lorentzian counterpart has been implemented by Kru\v{c}kovi\v{c} \cite{Kruchkovich:1954} and recently refined in 
\cite{Nozawa:2019dwu} by making use of obstruction criteria for KVs (for the underlying geometric analysis of these geometries, 
see  \cite{Chow:2019ucq}). This is another intriguing opportunity offered by 
recent developments in \cite{Kruglikov:2018qcn,Nozawa:2019dwu}. 

%% # of KVs with diff. invs. -> a class of (M, g): review 2 and the aim of this paper
The difficulty for the complete classification of local metrics tends to increase as the total number of KVs becomes fewer. 
A primary study in this line was due to Bianchi \cite{Bianchi:1898}, who elaborated on
the three-dimensional Riemannian manifolds with an isometry group of transitive action.   
The geometry of their metrics has been intensively studied in the literature \cite{Milnor:1976} since 
it has a prominent application to homogeneous cosmological models \cite{Ellis:1968vb}. 
Consequently, local forms of three-dimensional Riemannian metrics with 3 KVs have been known. 
However, the Lorentzian counterpart is still lacking in spite of the intensive results on left-invariant Lorentzian metrics on three-dimensional Lie groups \cite{Rahmani:1992,Cordero:1995,Cordero:1997}. Our aim in this paper is to present a complete catalogue of local Lorentzian metrics that admit three KVs.\footnote{The classification device employed by Petrov~\cite{Petrov} based on the canonical real structures has been successfully able to construct canonical form of metrics admitting KVs in four dimensional (pseudo-)Riemannian manifold. However, this method seems to call for excessive and brute-force computations. }
We would like to emphasize that not all the Lorentzian metrics with three KVs are recovered by a mere Wick-rotation of 
Riemannian metrics with three KVs. A principal reason is that the Lorentzian metrics may admit a null KV, which cannot be captured by 
complexification of the coordinates.   

Bona and Coll have obtained the necessary and sufficient conditions for the three-dimensional Lorentzian metrics 
to admit isometry group according to its isotropy subgroup,  in terms of eigenvalues and eigenvectors of the Ricci tensor \cite{BonaColl}. 
Nevertheless, the local form of the metrics has not been explored. To fill this gap is one of the main result in the present paper. 
It is also noteworthy to remark that our technique allows the complete classification without appealing to how the isometry group acts on the manifold. 

Our strategy for the classification of metrics admitting KVs is based on the obstructions for the existence of KVs, which amount to the 
algebraic constraints obeyed by curvature tensors and connections. 
Conceptually, this blueprint is closely parallel with the classification of supersymmetric solutions in supergravity \cite{Maeda:2011sh,Gran:2018ijr}. The existence of Killing spinors puts constraints on the intrinsic torsion of $G$ structures, which tightly restricts the possible form of metrics and fluxes. 
In our case, the role of intrinsic torsion is played by the two obstruction matrices described below.

Before stating our result, let us first evoke the definition and some key properties of KVs. 
Let  $(M, g)$  denote the three-dimensional Lorentzian manifold. 
A vector field $K^a$ on $(M, g)$  is a KV iff it satisfies the Killing equation
\begin{align}
\ma L_K g_{ab}~=~2\nabla_{(a} K_{b)}
~=~0 \,,
\label{eq:Killing}
\end{align}
where $\ma L_K$ is the Lie derivative along $K^a$, $\nabla_a$ denotes the Levi-Civita connection and
the round brackets denote symmetrization over the enclosed indices.
We raise and lower  the indices by $g_{ab}$ and its inverse $g^{ab}$. 
The compatibility of \eqref{eq:Killing} leads to the defining equation of a curvature collineation \cite{Kerr:1962}
\begin{align}
\ma L_K R_{abc}{}^d ~=~ 0\:,
\end{align}
which gives rise to the following zero eigenvalue problem
\begin{align}
\label{eq:first}
&(\bs R^{\mathrm{1st}})_a\:K^a ~=~ 0\,,
&&(\bs R^{\mathrm{1st}})_a ~\equiv~ \left(
\begin{array}{c}
\nabla _a R   \\
\nabla _a S^{(2)}  \\
\nabla _a S^{(3)} 
\end{array}
\right)\,,
\end{align}
where $R_{abc}{}^d$ is the Riemann--Christoffel tensor defined by $[\nabla_a, \nabla_b]V_c=R_{abc}{}^d V_d$ for any covector $V_a$.
A triple of curvature invariants $\{R, S^{(2)}, S^{(3)}\}$ are constructed as
$R \equiv R^a{}_a$, $S^{(2)} \equiv S^a{}_bS^b{}_a$ and $S^{(3)}\equiv  S^a{}_bS^b{}_cS^c{}_a$
with the Ricci tensor $R_{ab}\equiv R_{acb}{}^c$ and its traceless part $S_{ab} \equiv R_{ab}-(1/3)R g_{ab}$.
We call $\bs R^{\mathrm{1st}}$ the {\it first obstruction matrix},
since its minors put restrictions on the dimension of the solution space of \eqref{eq:Killing}.
A general solution of \eqref{eq:first} can be written as
\begin{align}
K^a ~=~ \sum_{\alpha=1}^d \omega_\alpha\:e_\alpha{}^a\:,
\label{eq:ansatz}
\end{align}
where $d \equiv \dim \ker \bs R^{\mathrm{1st}} \leq 3$, $\{e_\alpha{}^a\}$ are linearly independent vectors that span $\ker \bs R^{\mathrm{1st}}$
and $\{\omega_\alpha\}$ are arbitrary functions on $M$.
Substituting equation \eqref{eq:ansatz} into \eqref{eq:Killing}, we end up with a PDE system of the form
\begin{align}
&\nabla_a {\boldsymbol \omega} ~=~ {\boldsymbol \Omega}_a\:{\boldsymbol \omega}\:,
&&{\boldsymbol \omega} ~\equiv~
\left(
\begin{array}{c}
\omega_\alpha \\
\omega_{\alpha \beta} \\
\end{array}
\right)\,,
\label{eq:PDE}
\end{align}
where $\{\omega_{\alpha \beta} \equiv \ma L_{[\alpha} \omega_{\beta]} \}$ are the 1-jet variables and
the abbreviation $\ma L_\alpha$ denotes the Lie derivative along $e_\alpha{}^a$.
The square brackets over indices is used for skew-symmetrization.
The connection ${\boldsymbol \Omega}_a$ is expressed in terms of  the Ricci rotation coefficients, curvatures and their derivatives.
Since the system \eqref{eq:PDE} is first-order, 
its compatibility yields a set of algebraic equations
\begin{align}
&(\bs R^{\mathrm{2nd}})\:\bs \omega ~=~0 \,,
&& (\bs R^{\mathrm{2nd}})_{ab} ~\equiv~  \nabla_{[a}\bs \Omega_{b]} -\bs \Omega_{[a}\bs \Omega_{b]} \,. 
\label{eq:second}
\end{align}
We call $\bs R^{\mathrm{2nd}}$ the {\it second obstruction matrix}, whose entries are also obstructions to local existence of KVs.
Note that the matrix depends upon maximal possible dimension of an orbit space under an isometric group action, and of its isotropy subgroup.
At the outset, they are  given by $d = \dim \ker \bs R^{\mathrm{1st}}$ and $d(d-1)/2$, respectively.
By deriving a complete list of the second obstruction matrices, all the conditions to ensure that the manifold $M$ attains three KVs have already been derived in \cite{Nozawa:2019dwu}. 
Turning the logic around, these obstructions can be used to obtain the explicit local metrics with three KVs. 
Our principal result of this paper can be boiled down as follows:

\bigskip \noindent \textbf{Theorem}
	{\it 
	Let $(M, g)$ be a three-dimensional Lorentzian manifold.
	If it admits exactly three linearly independent Killing vectors,
	the manifold is locally isometric to any one of the metrics described in table~\ref{tab:theorem}.}

\begin{table}[htb]
	\caption{Normal forms of homogeneous metrics with three linearly independent KVs. The metrics in referred equations in this table 
		are $-(e_1)^2+(e_2)^2+(e_3)^2$.
		$\D \Sigma_k^2$ describes the two-dimensional maximally symmetric space with a constant curvature $k=0, \pm 1$, and 
		$\D \ti \Sigma_k^2$ corresponds to its Lorentzian counterpart.
		The quantity $\sigma$ is constant in this table and is defined by (\ref{eq:sigma}). $\lambda $, $C_0, C_1, C_2$ are also constants. 
		For the Ricci rotation coefficients $\{\kappa_i, \eta_i, \tau_i\}$ $(i=1,2,3$ or $u, v, e)$ which are all constants in this table, 
		we refer the reader to (\ref{eq:inhom1:RRC}) and (\ref{null:RRC}).
	}
	\begin{tabular}{c|c|c|c|c} \hline \hline
		Class & Segre type& \multicolumn{2}{|c|}{Branch} & Canonical metric  \\ \hline
		\multirow{3}{*}{Inhomogeneous}  & $[1,(11)] $ &\multicolumn{2}{|c|}{--} & $-\D t^2+a^2 (t)\D \Sigma_k^2$  \\
		& $[(1,1)1] $ & \multicolumn{2}{|c|}{--} &$\D t^2+a^2 (t)\D \ti \Sigma_k^2$  \\ 
		& $[(21)]$&\multicolumn{2}{|c|}{--} & $e^{2f(x)} (2\D x\D y+\D z^2 )$  \\  \hline 
		\multirow{20}{*}{Homogeneous}  & \multirow{5}{*}{$[1,(11)]$}  & & $\tau_2=0$  & (\ref{[1,(11)]metric:tau2+tau3=0:tau2=0})  \\
		& & $\tau_2+\tau_3=0$ &  $\tau_2^2-\kappa_2^2<0 $ & (\ref{[1,(11)]metric:tau2+tau3=0:tau2sq-kappa2sq<0}) 
		\\
		&  & & $\tau_2^2-\kappa_2^2>0 $  & (\ref{[1,(11)]metric:tau2+tau3=0:tau2sq-kappa2sq>0})    \\ \cline{3-5}
		& &\multirow{2}{*}{$\tau_2+\tau_3\ne0 $}& $\tau_2 \tau_3>0$ & (\ref{[1,(11)]metric:tau_2+tau3ne0:tau2tau3>0}) \\ 
		& & & $\tau_2 \tau_3<0$  & (\ref{[1,(11)]metric:tau_2+tau3ne0:tau2tau3<0}) \\ \cline{2-5}
		& \multirow{4}{*}{$[(1,1)1]$}   &\multirow{2}{*}{$\tau_1=\tau_3 $} & $\tau_1=0$ & (\ref{[(1,1)1]metric:tau1=tau3:tau1=0}) \\ 
		& & & $\tau_1 \ne 0$ & (\ref{[(1,1)1]metric:tau1=tau3:tau1ne0}) \\ \cline{3-5}
		& & \multirow{2}{*}{$\tau_1\ne \tau_3 $} & $4\kappa_1^2\ne (\tau_1-\tau_3)^2$ & 
		(\ref{[(1,1)1]metric:tau1netau3:4kappa1sqne(tau1-tau3)sq})  \\
		& & &$4\kappa_1^2=(\tau_1-\tau_3)^2$ & (\ref{[(1,1)1]metric:tau1netau3:4kappa1sq=(tau1-tau3)sq})     \\ \cline{2-5}
		& \multirow{3}{*}{$[1,11]$} & \multicolumn{2}{|c|}{$\kappa_2=\eta_1=\eta_3=0$}  & (\ref{[1,11]metric:(A-i-1)})   \\ \cline{3-5} 
		& &\multicolumn{2}{|c|}{$\kappa_2=\eta_3=\tau_2-\tau_1=0$ with $\eta_1\ne 0$}   & (\ref{[1,11]metric:(A-ii)})  \\ \cline{3-5} 
		& &\multicolumn{2}{|c|}{$\eta_3=\tau_3-\tau_2=\eta_1=0$ with $\kappa_2 \ne 0$}  & (\ref{[1,11]metric:(B-i-1)})  \\ \cline{2-5}
		& $[( 2,1)]$ & \multicolumn{2}{|c|}{$\sigma \ne 0$} &  $2 e^{-2\tau_v z}\D x \D y+ \D  z^2 +C_0 e^{2(\sigma-2\tau_v)z}\D x^2 $  \\ \cline{2-5}
		& \multirow{3}{*}{$[2,1]$}   & \multicolumn{2}{|c|}{$\tau_u+\tau_v \ne 0 $} & 
		$2 [\D x+2(x-3\lambda y)\tau_v \D z ] \left(\D y-\frac{x}{4\tau_v} \D z\right)+\D z^2$ 
		\\    \cline{3-5}
		& &\multirow{2}{*}{$\tau_u+\tau_v =  0$}   &$\sigma= 2\tau_v$   &
		$2 \D x \left[\D y-\left(C _1z +\sqrt{2\tau_v C_1}\Sigma y \right)\D x\right]+(\D z+2\tau_vy \D x)^2$ \\   \cline{4-5}
		& & & $\sigma_*\equiv \sigma-2\tau_v \ne 0$  &
		$2\D x \left[\D y +\left(-C_2 e^{2\sigma_* z} + \sigma_* \tau_v y^2 \right)\D x \right]+(\D z+2 \tau_v y \D x)^2$
		\\  \cline{2-5}
		& $[3]$ & \multicolumn{2}{|c|}{--} & 
		$2 \D x \left[e^{-2\tau_v z} \D y - \left(\frac{y}{2\tau_v}+\frac{z}{8\tau_v^3}\right)e^{-4\tau_v z}\D x\right]+\D z^2$  \\ \cline{2-5}
		& [$z\bar z1$] &\multicolumn{2}{|c|}{--} & 	
$-\left[\D t +\left(-\eta_1 t +\tau_3 x\right)\D y\right]^2+\D y^2+\left[\D x +\left(\eta_1x +\tau_3 t\right)\D y\right]^2$
	\\ \hline\hline
	\end{tabular}
	\label{tab:theorem}
\end{table}

\bigskip
An inspection of the explicit second obstruction matrices in \cite{Nozawa:2019dwu} shows that 
the three KVs appear for ``class 2'' and ``class 3'' in the terminology therein. 
We shall moniker class 2 as ``inhomogeneous'' and class 3 as ``homogeneous'' in this paper, 
since this epithet represents how the isometry group acts on the manifold.  
The classification of class 3 in \cite{Nozawa:2019dwu} relies on the property of 
Segre classification for the traceless Ricci tensor $S^a{}_b$.\footnote{
For a classification of $2+1$ dimensional asymptotically AdS spacetimes based on Segre types,  
see \cite{Glorioso:2015vrc}. } 
Note also that our classification of metrics with three KVs does not represent the existence of the global isometry group, 
which may be broken by the discrete identification of spacetime points. 
We refer the reader to e.g., \cite{Sheikh-Jabbari:2014nya} for the discussion of related aspects.

We note that in the ``homogeneous'' case  the Lie algebra and the local metric do not have a one-to-one correspondence. This is in sharp contrast to the Riemmanian case. 
For instance, metrics (\ref{[(1,1)1]metric:tau1=tau3:tau1ne0}), (\ref{metric[21]_caseBsigmanetauv}), (\ref{metric_[3]}) admit $\mathfrak{so}(1,2)$ algebra. The metric (\ref{[(1,1)1]metric:tau1=tau3:tau1ne0})  deserves a Lorentzian counterpart of Bianchi IX metric, while the latter two metrics are not. This illustrates the richness of Lorentzian signature. Therefore, we avoided to refer to the Bianchi types  throughout the text to circumvent the confusion. In addition, each Bianchi class counterpart is not realized as the three dimensional metrics with three KVs, since we are focusing on the three dimensional intrinsic metric, rather than the three dimensional subspace. 

Our result obtained in this paper may be generalized into higher dimensions, if we properly find out the obstruction matrices. 
Research in this direction is an intriguing future study. On top of this, 
some of the metrics (\ref{metric_[3]}), (\ref{metric_zz*1}) derived in the present paper appear to be new, as far as we know. 
It seems interesting to further explore physical aspects of these spacetimes.

The remainder of this paper is devoted to the proof of this theorem and is organized as follows. 
The next section \ref{inhomogeneous} classifies all the inhomogeneous metrics with three KVs, which turn out to be all 
conformally flat. The following seven sections \ref{sec_[1,(11)]},  \ref{sec_[(1,1)1]}, \ref{sec:[1,11]}, \ref{sec_Segre [(21)]}, 
\ref{sec_Segre[21]}, \ref{sec_Segre [3]}, \ref{sec_Segrezz*1} aim to the classification of homogeneous metrics
for each Segre type. Useful formulae for curvature tensors are summarized in appendix \ref{sec:app}.

\section{Inhomogeneous metrics}
\label{inhomogeneous}

This section identifies all the $3$-dimensional Lorentzian metrics endowed with an isometry group of dimension $3$ with
$1$-dimensional isotropy subgroup, acting on $2$-dimensional orbits.
Hereon, any KV can be written as
\begin{align}
K^a ~=~ \omega_2 \:e_2{}^a + \omega_3\:e_3{}^a\:,
\end{align}
where $\{e_2, e_3 \}$ are two annihilators of $\bs R^{\mathrm{1st}}$.
Depending on the causal character of annihilators, we divide the following analysis into three cases:
Subsection \ref{subsec:inhom1} treats the case in which the annihilators are both spacelike, 
Subsection \ref{subsec:inhom2} deals with the case one is spacelike while the other is timelike, and lastly 
Subsection \ref{subsec:null_and_spacelike} handles the case one of them is null. 
As it turns out, all metrics falling into this category are conformally flat with a conformal factor
depending only on a single variable. The three subclasses are characterized by the dependence on the 
timelike/spacelike/null coordinate. 
For the definition and useful formulae for these quantities, we refer the reader to appendix~\ref{sec:app}.

\subsection{Both annihilators are spacelike}
\label{subsec:inhom1}

We suppose in this subsection that an orthonormal frame $\{e_i\}$ $ (i = 1,2,3)$ is assigned to each point of $M$ as
\begin{align}
\label{eq:inhom1:g_inv}
g^{ab} ~=~ -e_1{}^a e_1{}^b +e_2{}^a e_2{}^b +e_3{}^a e_3{}^b\:.
\end{align}
By defining the Ricci rotation coefficients $\{\kappa_i, \eta_i, \tau_i\}$ as
(\ref{eq:inhom1:RRC}), 
the criterion for $(M, g)$ to have three linearly independent KV reads
(this can be read off from (3.6), (3.9) and (3.12) in \cite{Nozawa:2019dwu})
\begin{subequations}
	\label{eq:inhom1:criterion}
\begin{align}
\label{eq:inhom1:criterion_1}
0&~=~ {\kappa_2-\kappa_3}\:,\\
\label{eq:inhom1:criterion_2}
0&~=~\kappa_1 ~=~ \eta_1 ~=~ \tau_2 ~=~ \tau_3\:,\\
\label{eq:inhom1:criterion_3}
0&~=~\ma L_2 \kappa_2 ~=~ \ma L_3 \kappa_2\:,\\
\label{eq:inhom1:criterion_4}
0&~=~\ma L_2 R_{22} ~=~ \ma L_3 R_{22}\:,
\end{align}
\end{subequations}
where $\ma L_i$ is the Lie derivative along $e_i{}^a$ and $R_{ij} \equiv R_{ab} e_i{}^ae_j{}^b$ $(i,j=1,2,3)$.
It immediately follows from \eqref{eq:inhom1:criterion_1}, \eqref{eq:inhom1:criterion_2} and (\ref{RRC_nonull}) that 
\begin{align}
\label{}
&\nabla_{[a} e_{1b]} ~=~0 \,,
&&W_{[a}\nabla_{b }W_{c]} ~=~0 \,, 
\end{align}
where $W_a\equiv e_{2a}+i e_{3a}$ is a complex vector.
Then, there exists a local coordinate chart $(t,x,y)$ such that
\begin{align}
\label{}
&e_{1a} ~=~- \nabla_a t \,,
&&W_a ~=~ e^{i \theta _1(t,x,y)+\theta_2(t,x,y)}(\nabla _a x+i \nabla_a y) \,, 
\end{align}
where $\theta_1$ and $\theta_2$ are real functions.
It is easy to see that the following gauge transformation with a parameter $\lambda$
\begin{align}
\label{eq:imhom1:gauge}
&e_2{}^a ~\to~ (\cos \lambda)\: e_2{}^a +(\sin \lambda)\:e_3{}^a \:,
&&e_3{}^a ~\to~ -(\sin \lambda)\:e_2{}^a + (\cos \lambda)\:e_3{}^a\:,
\end{align}
leaves \eqref{eq:inhom1:criterion} invariant.\footnote{
We illustrate in Appendix \ref{sec:app:ortho} how the Ricci rotation coefficients \eqref{eq:inhom1:RRC} are transformed under the transformation \eqref{eq:imhom1:gauge}.}
One can exploit this gauge freedom to set $\theta_1 = 0$ without loss of generality.
Equations \eqref{eq:inhom1:criterion_3} and \eqref{eq:inhom1:criterion_4} then boil down to
\begin{align}
\label{}
\theta_2(t,x,y) ~=~ \log a(t)+ \psi (x,y)\:,
\end{align}
with 
\begin{align}
\label{Liouvilleeq}
(  \partial_x^2+\partial_y^2 ) \psi ~=~ -k\:e^{2\psi}\:,
\end{align}
where $a$ and $\psi$ are arbitrary functions and $k$ is a constant which can be normalized to be $0$ or $\pm 1$.
The {equation} \eqref{Liouvilleeq} stands for Liouville's equation on a surface of constant Gaussian curvature $k$,
whose metric takes the form $\D \Sigma_k^2 \equiv e^{2\psi(x,y)}(\D x^2+\D y^2)$.
One finds that \eqref{eq:inhom1:g_inv} becomes
\begin{align}
\label{eq:inhom1:result}
\D s^2~=~- \D t^2+ a^2(t)\D \Sigma_k^2 \,. 
\end{align}
This metric describes the 2+1 dimensional Friedmann-Lema\^itre-Robertson-Walker (FLRW) universe.
Note that the set of all isometries of $\D \Sigma_k^2$ forms the isometry group of the whole spacetime \eqref{eq:inhom1:result}.

Now, let us consider the metric \eqref{eq:inhom1:result} together with a particular solution to \eqref{Liouvilleeq} and its isometry algebra.
A solution to Liouville's equation \eqref{Liouvilleeq} can be taken as
\begin{align}
\label{Liouvillesol}
\psi ~=~ -\log \left[ 1+\frac k4(x^2+y^2)\right]\,.
\end{align}
Performing the change of variables $(x, y) \to (\theta, \phi)$,
given by the relation $x+i y =\tfrac{2}{\sqrt k}e^{i\phi} \tan \left(\tfrac{\sqrt k}{2}r\right)$,  
the metric reads $\D \Sigma_k^2 = \D \theta^2+[\tfrac 1{\sqrt k} \sin (\sqrt k r) ]^2 \D \phi^2$. % for short.
Then, the three linearly independent KVs can be found in the form
\begin{align}
\label{}
&K_1 ~=~ \sin\phi \:\partial_r +\sqrt k \cos\phi \cot (\sqrt k r )\:\partial_\phi \,,
&&K_2~=~ \cos\phi\:\partial_r -\sqrt k \cot (\sqrt k r)\sin\phi \:\partial_\phi \,,
&&K_3 ~=~ \partial_\phi \,,
\end{align}
with the commutation relations
\begin{align}
\label{eq:inhom1:algebra}
&[K_1, K_2]~=~k K_3 \,,
&&[K_2, K_3] ~=~ K_1 \,,
&&[K_3, K_1] ~=~ K_ 2 \,.
\end{align}
This algebra corresponds to $\mathfrak{so}(3) $ for $k=1$, $\mathfrak e^{2} $ for $k=0$ and 
$\mathfrak{so}(2,1)$ for $k=-1$.

\subsection{One annihilator is timelike}
\label{subsec:inhom2}

This subsection is essentially a duplication of subsection \ref{subsec:inhom1}.
As in the previous subsection, we assume that $\{e_i\}$ $(i=1,2,3)$ forms an orthonormal frame
\begin{align}
\label{eq:inhom2:g_inv}
g^{ab} ~=~ e_1{}^a e_1{}^b +e_2{}^a e_2{}^b -e_3{}^a e_3{}^b\:.
\end{align}
In terms of the Ricci rotation coefficients $\{\kappa_i, \eta_i, \tau_i\}$ defined in \eqref{eq:inhom1:RRC},
the requirement that the manifold $(M, g)$ admits exactly three linearly independent KVs can be written as 
{\eqref{eq:inhom1:criterion_2}--\eqref{eq:inhom1:criterion_4} and
$\kappa_2+\kappa_3 =0$.}
It follows that 
\begin{align}
\label{}
&\nabla_{[a} e_{1b]} ~=~0 \:,
&&W^\pm _{[a}\nabla_b W^\pm_{c]} ~=~0 \:,
\end{align}
where $W^\pm _a \equiv e_{2a}\pm e_{3a} $ denote two real null vectors.
Thereupon, a local coordinate chart $(t,x,y)$ can be chosen as
\begin{align}
\label{}
&e_{1a} ~=~\nabla_a t \,,
&&W^+_a ~=~ e^{\theta_1 (t,x,y)} \nabla_a x \,,
&&W^- _a ~=~ e^{\theta_2 (t,x,y)} \nabla_a y \,.
\end{align}
Since there exists a gauge transformation with an arbitrary function $\lambda$
\begin{align}
\label{eq:imhom2:gauge}
&e_2{}^a ~\to~ (\cosh \lambda)\: e_2{}^a+(\sinh \lambda)\:e_3{}^a \:,
&&e_3{}^a ~\to~ (\sinh \lambda)\:e_2{}^a + (\cosh \lambda)\:e_3{}^a\:,
\end{align}
which enables us to choose $\theta_1=\theta_2\equiv \theta$. 
The remaining equations  yield
\begin{align}
\label{eq:inhom2:Liouville}
\theta(t,x,y) ~=~\log a (t)+\psi (x,y)\:,
\end{align}
with 
\begin{align}
\label{null_Liouvilleeq}
\partial_x\partial_y \psi ~=~-\frac k4 \:e^{2\psi}\:, 
\end{align}
where $a$ and $\psi$ are arbitrary functions. 
The PDE for $\psi$ describes a Liouville surface equipped with the metric $\D \tilde \Sigma^2_k= e^{2\psi}\D x \D y$ and corresponding Gaussian curvature $k =0, \pm 1$.
The metric then reads 
\begin{align}
\label{eq:inhom2:result}
\D s^2 =~ \D t^2+ a^2(t)\D \tilde \Sigma_k^2 \,.
\end{align}
This is the double Wick-rotated metric of \eqref{eq:inhom1:result}, as expected.

{For further consideration, let us take a particular solution to \eqref{null_Liouvilleeq} of the form}
\begin{align}
\label{null_Liouvillesol}
\psi ~=~- \log \left(1+\frac k4 xy \right) \,,
\end{align}
and thereby the local metric of $\D \ti \Sigma_k^2$ is then represented as
$\D \ti \Sigma_k^2 = \D \theta^2 - [\tfrac 1{\sqrt k} \sin (\sqrt k r) ]^2 \D \phi^2$, 
together with new variables ($r, \phi $) defined by $x=\tfrac{2}{\sqrt k}\tan \left(\tfrac{\sqrt k}{2}r \right)e^{\phi}$ and
$y=\tfrac{2}{\sqrt k}\tan \left(\tfrac{\sqrt k}{2}r\right)e^{-\phi}$.
Now, the three KVs are given by
\begin{align}
\label{EucFLRW_KVs}
&K_1 ~=~ -\sinh \phi\:\partial_r +\sqrt k \cosh\phi \cot (\sqrt k r)\:\partial_\phi \,,
&&K_2 ~=~ \cosh \phi\:\partial_r -\sqrt k \sinh \phi \cot (\sqrt kr )\:\partial_\phi \,,
&&K_3 ~=~  \partial_\phi \,, 
\end{align}
entailing the commutators
\begin{align}
\label{EucFLRW_algebra}
&[K_1, K_2]~=~k K_3 \,,
&&[K_2, K_3] ~=~K_1 \,,
&&[K_3, K_1]~=~- K_2 \,.
\end{align}
This defines $\mathfrak{so}(1,2)$ for $k=\pm 1$ and 
Poincar\'e algebra $\mathfrak e^{1,1}$ for $k=0$.

\subsection{One annihilator is null}
\label{subsec:null_and_spacelike}

Let us consider the case in which one of the annihilators is null. One can write the KV as 
\begin{align}
\label{KV_class2_null}
K^a ~=~ \omega_u\:u^a +\omega_e\:e^a \,, 
\end{align}
where we have adopted a null frame $\{u,v,e \}$
\begin{align}
\label{}
g^{ab} ~=~2u^{(a}v^{b)} +e^a e^b \,,
\end{align}
that satisfies
\begin{align}
&u^av_a ~=~e^ae_a ~=~1 \,,
&&u^au_a ~=~ v^av_a ~=~ u^ae_a ~=~ v^a e_a ~=~ 0\:.
\end{align}
In this case, the 3 KV can arise when $u^a$ is a geodesic tangent
and  the 1-jet variable $\varpi \equiv \ma L_v \omega_e$ is functionally independent of $\omega_u$ and $\omega_e$
\cite{Nozawa:2019dwu}.
In terms of the Ricci rotation coefficients $\{\kappa_i, \eta_i, \tau_i\}$ $(i=u,v,e)$ defined in \eqref{null:RRC}
and frame components of the Ricci tensor \eqref{eq:app_ricci_null}, this only occurs for 
(see eqs. (3.19), (3.22) and (3.25) in \cite{Nozawa:2019dwu})
\begin{subequations}
\begin{align}
\label{class2_null_branch}
\eta_u ~=~ \tau_v ~=~\kappa_e~=~0 \,,
\end{align}
and
\begin{align}
\label{class2_null_2ndobs}
&R_{uv}~=~R_{ve}~=~0 \,,
&&\ma L_e R_{vv}+2\tau_e  R_{vv}~=~0 \,.
\end{align}
\end{subequations}

The KV in the form (\ref{KV_class2_null}) allows two kinds of frame transformations
(\ref{null_localSO21_I}) and (\ref{null_localSO21_II}). 
One can exploit this gauge freedom (\ref{null_localSO21_I}) to obtain 
$\tau_u+\tau_e=0$, while the other conditions  (\ref{class2_null_branch}), 
(\ref{class2_null_2ndobs}) remain inert under (\ref{null_localSO21_I}). 
Since $u^a $ and $e^a $ are now commutative (c.f., (\ref{eq:app_comm_null})), we can work in the coordinate system 
($x,y,z$) such that 
\begin{align}
\label{nullframe_uecommute}
&u^a ~=~ (\partial_y )^a \,,
&&v^a ~=~ V_1(\partial_x)^a 
+V_2(\partial_y)^a +V_3 (\partial_z)^a  \,,
&&e^a ~=~ (\partial_z)^a \,,
\end{align}
where $V_i$ are functions of $x,y,z$. Lowering indices, we have
\begin{align}
\label{}
&u_a ~=~ \frac 1{V_1} \nabla_a x \,,
&&v_a ~=~ \nabla_a y -\frac{V_2}{V_1} \nabla_a x \,,
&&e_a ~=~ \nabla_a z -\frac{V_3}{V_1} \nabla_a x \,.
\end{align}
In this coordinate system, $\tau_v=0$ is solved as 
\begin{align}
\label{null_V13}
&V_1 ~=~\frac{1}{\partial_y F} \,,
&&V_3 ~=~-\frac{\partial_zF}{\partial_y F} \,,
\end{align}
where $F=F(x,y,z)$. 
The first two conditions in \eqref{class2_null_2ndobs} are twice integrated to yield 
\begin{align}
\label{null_V2}
V_2 ~=~ -\frac{1}{2(\partial_yF)^2}[2Ff_1+f_2 -(\partial_zF)^2+2 \partial_x F] \,, 
\end{align}
where $f_1=f_1(x)$ and $f_2=f_2(x,z)$. 
The last condition in \eqref{class2_null_2ndobs} is satisfied, provided that $\partial_z^3 f_2=0$ holds.
This gives
\begin{align}
\label{}
f_2(x,z)~=~f_{20}(x)+f_{21}(x)z+f_{22}(x)z^2 \,. 
\end{align}
Changing variables to $\hat x=h(x)$, $\ti y=F(x,y,z)/h'(x)$ and 
choosing $h''(x)=-f_1(x)h'(x)$, we obtain
\begin{align}
\label{}
\D s^2=2 \D\ti  y \D \hat x+\D \hat x^2[
\hat f_{20}(\hat x)+\hat f_{21}(\hat x)z+\hat f_{22}(\hat x)z^2]+\D z^2 \,, 
\end{align}
where $\hat f_{2i}(\hat x)=h'(x)^{-2} f_{2i}(x)$. The functions $\hat f_{20}$ and $\hat f_{21}$ can be made to vanish 
by the transformation $z=\hat z+h_1(\hat x)$ and $\ti y=\hat y+h_2(\hat x)-h_1'(\hat x) \hat z$, thus giving
\begin{align}
\label{}
\D s^2 ~=~ 2 \D \hat x \D \hat y+\D \hat z^2+\hat z^2 \hat f_{22}(\hat x) \D\hat x^2 \,, 
\end{align}
where we have chosen $h_1, h_2 $ to satisfy $\hat f_{21}+2\hat f_{22}h_1-2h_1''=0$ and 
$\hat f_{20}+\hat f_{21}h_1+\hat f_{22}h_1^2+h_1'^2+2h_2'=0$. 
One can verify that the Cotton tensor for this metric vanishes, meaning that the metric is conformally flat. 
To render the conformal flatness manifest, we perform 
further coordinate transformations $\hat x=\int e^{2f(x)}\D x$, 
$\hat z=e^{f(x)} z$, $\hat y=y-\frac 12 z^2 f'(x)$ 
with $e^{4f}\hat f_{22}+f'^2-f''=0$, which brings the metric into 
\begin{align}
\label{}
\D s^2 ~=~ e^{2f(x)} (2\D x\D y+\D z^2) \,,
\end{align}
where $f(x)$ should meet the condition $f'' \ne f'^2$.

The three KVs read
\begin{align}
\label{}
&K_1~=~-z\partial_y +x \partial_z \,,
&&K_2~=~\partial_y \,,
&&K_3~=~\partial_z \,, 
\end{align}
whose nonvanishing commutators are
\begin{align}
\label{}
[K_1, K_3]=K_2 \,. 
\end{align}
This defines the Heisenberg algebra. 
Since $K_2$ is covariantly constant null vector ($\nabla_a K_{2b}=0$, $K_{2a}K^a_2=0$), 
the metric describes the pp-wave spacetime.

\section{Homogeneous metrics: Segre type [1,(11)]}
\label{sec_[1,(11)]}

Let us next move on to the homogeneous case, for which the first obstruction matrix trivially vanishes. 
As discussed in our previous paper \cite{Nozawa:2019dwu}, one can address this case by resorting to the Jordan decomposition of the 
trace-free Ricci tensor $S^a{}_b$, i.e., Segre type. When the metric is given, the transformation to the Jordan basis requires us to solve 
the eigenvalue problem, which is a prime impediment in practice.  
In contrast, 
the Jordan basis is of great help in reducing total amount of computations, 
as far as the classification of the metrics with KVs is concerned, 

Let us begin with our discussion for the the Segre [1,(11)] type, for which the trace-free Ricci tensor 
takes the following form 
\begin{align}
\label{Segre[1,(11)]}
S_{ab}=-\lambda (2 e_{1a}e_{1b}+e_{2a}e_{2b}+e_{3a}e_{3b})\,, 
\end{align}
where $\{e_i \}$ is the orthonormal frame (\ref{nonnull_frame}) with $\varepsilon=-1$. 
Class 3 condition in \cite{Nozawa:2019dwu} requires that any KV takes the form 
\begin{align}
\label{}
K^a =\omega_1 e_{1a} +\omega_2 e_{2a}+\omega_3 e_{3a} \,, 
\end{align}
and the curvature components must be subjected to
\begin{align}
\label{}
\{\lambda, R\} ={\rm constants.} 
\end{align}
From the Bianchi identity and the invariance condition $\ma L_K S_{ab}=0 $, we have 
\begin{align}
\label{Segre[1,(11)]_Bianchi}
\kappa_1=0 \,, \qquad \eta_1=0 \,, \qquad \kappa_3=-\kappa_2 \,. 
\end{align}
For the case $\kappa_2=\tau_2+\tau_3=0$, we have 4 KVs which we shall not discuss. 
The 3 KVs may appear in the two cases below, which are distinguished 
according to $\tau_2+\tau_3=0 $ or not.

\subsection{$\tau_2+\tau_3=0 $}

When 
\begin{align}
\label{}
\tau_2+\tau_3=0 \,, \qquad \kappa_2 \ne 0 \,, 
\end{align}
the 1-jet variable $\varpi_3 =\ma L_2 \omega_3$ is linearly dependent on $\{\omega_i\}$ and 
the second obstruction matrix is given by (4.12c) in~\cite{Nozawa:2019dwu}, 
which vanishes iff 
\begin{align}
\label{}
\{\tau_2, \eta_2 ,\eta_3\} ={\rm constants.} 
\end{align}
Here $\ma L_1 \tau_2=0$ follows from $R_{23}=R_{32}$. 
Segre type [1,(11)] requires the constancy of 
$R_{11}=2(\tau_2^2-\kappa_2^2)$,  implying that $\kappa_2 (\ne 0)$ 
is also a constant. 
Since $S_{12}=S_{13}=S_{23}=0$ must be fulfilled in the Jordan basis of type [1,(11)], 
we obtain $\eta_2=\eta_3 =\tau_1=0$. 
It follows that only the nonvanishing Ricci rotation coefficients are $\kappa_2$ and $\tau_2$, both of 
which are constants parameterizing the homogeneous solution. The curvature tensors now reduce to 
\begin{align}
\label{}
\lambda =\frac 23 (\kappa_2^2-\tau_2^2) \ne 0 \,, \qquad 
R= 2(\kappa_2^2-\tau_2^2) \,. 
\end{align}
Plugging these expressions into Einstein's equations, these correspond to the cosmological constant and the 
energy density of the dust fluid. 

Then, the orthonormal frame $e_i$ obeys the following first-order relations
\begin{subequations}
\begin{align}
\label{}
\nabla_b e_{1a}=&\,  \kappa_2 (-e_{2b}e_{2a}+e_{3b}e_{3a})+\tau_2( e_{2b} e_{3a}-e_{3b}e_{2a}) 
\,,   \\ 
\nabla_b e_{2a}=&\, -\kappa_2 e_{2b}e_{1a}-\tau_2 e_{3b}e_{1a} \,, \\
\nabla_{b} e_{3a}=&\,  \tau_2e_{2b}  e_{1a}{+\kappa_2}e_{3b}e_{1a}\,. 
\end{align}
\end{subequations} 
It turns out that $W_a =e_{2a}+\alpha e_{3a}$ satisfies 
\begin{align}
\label{}
\nabla_b W_a =(-\kappa_2+\alpha \tau_2 )e_{1a} W_b \,, 
\end{align}
where $\alpha $ is the solution to the quadratic equation
\begin{align}
\label{[1,(11)]_alphaeq}
\tau_2 \alpha^2-2\kappa_2 \alpha +\tau_2 =0 \,. 
\end{align}
Depending on the nature of roots for this equation, 
we can obtain individual local metrics. We remark 
that the none of the above conditions imposed on Ricci rotation coefficients 
are invariant under further transformations (\ref{nonnull_localSO21}). Specifically, 
the triad frame has been already fixed completely.

\subsubsection{$\tau_2=0$}

Let us first consider the $\tau_2=0$ case. Since $e_{i}$ are all hypersurface-orthogonal, 
we can introduce local coordinates by
\begin{align}
\label{}
e_{1a}= -e^{\phi_0 (t,x,y)} \D t \,, \qquad 
e_{2a}= e^{\phi_1 (t,x,y)} \D x \,, \qquad 
e_{3a}= e^{\phi_2 (t,x,y)} \D y \,. 
\end{align}
$\kappa_1=\eta_1=0$ implies that $\partial_x \phi_0=\partial_y \phi_0=0$, 
so that we can set $\phi_0=0$ by the redefinition  of $t$. Conditions
$\kappa_2+\kappa_3=\eta_2=\eta_3=0$ are now solved to give 
$\phi_1(t,x,y)=\log a(t)+\Phi_{1}(x)$ and  $\phi_2(t,x,y)=-\log a(t)+\Phi_{2}(y)$, 
thence we can set $\Phi_1=\Phi_2=0$ by absorbing into the definition of $x$ and $y$. Lastly, $\kappa_2={\rm constant}$ gives 
$a(t)=e^{-\kappa_2 t}$ and the metric belongs to the ($2+1$)-dimensional Bianchi I class
\begin{align}
\label{[1,(11)]metric:tau2+tau3=0:tau2=0}
\D s^2= -\D t^2+ e^{-2\kappa_2 t}\D x^2+ e^{2\kappa_2 t } \D y^2 \,. 
\end{align}
The Killing vectors are given by
\begin{align}
\label{}
K_1=\partial_x \,, \qquad 
K_2=\partial_y \,, \qquad 
K_3=\frac 1{\kappa_2}\partial_t +x \partial_x -y \partial_y \,. 
\end{align}
The nonvanishing commutation relations are  
\begin{align}
\label{}
[K_2,K_3 ]=-K_2 \,, \qquad [K_3, K_1]=-K_1 \,. 
\end{align}
This corresponds to $\mathfrak e^{1,1}$ algebra.

\subsubsection{$\tau_2^2-\kappa_2^2<0$ and $\tau_2\ne 0$}

Since the equation (\ref{[1,(11)]_alphaeq}) admits two real roots 
$\alpha_\pm =\tau_2^{-1}(\kappa_2 \pm \sqrt{\kappa_2^2-\tau_2^2})$, 
it follows that $W^\pm _a = e_{2a}+\alpha_\pm e_{3a} $ satisfy
\begin{align}
\label{}
\nabla_b W^\pm _a =\pm \sqrt{\kappa_2^2-\tau_2^2 } e_{1a} W^\pm _b \,, \qquad 
g^{ab}W^+_a W^-_b =2 \,. 
\end{align}
One can then introduce coordinates by
\begin{align}
\label{}
e_{1a}=- f(\nabla_a t+\chi_1 \nabla_a  x+\chi_2  \nabla_a  y )\,, \qquad 
W^+_a= e^{\phi_+ }  \nabla_a x \,, \qquad 
W^-_a= e^{\phi_- }  \nabla_a y \,, 
\end{align}
where $f=f(t,x,y)$, $\chi_{1,2}=\chi_{1,2}(t,x,y)$ and $\phi_\pm =\phi_\pm (t,x,y)$. 
By $t\to \int f^{-1}\D t$, one can achieve $f\equiv 1$ modulo the 
redefinition of $\chi_{1,2}$. 
From $\kappa_2=-\kappa_3={\rm const}$, one obtains 
$\phi_-+\phi_+=2 \psi (x,y)$ and 
$\phi_+=t \sqrt{\kappa_2^2-\tau_2^2}+\Psi(x,y)$. 
The condition $\kappa_1=\eta_1=0$ implies 
$\partial_t \chi_{1,2}=0$, whereas $\eta_3=0 $ gives 
\begin{align}
\label{}
\chi_2=\frac{\partial_y \Psi}{\sqrt{\kappa_2^2-\tau_2^2}}\,, \qquad 
\chi_1=\frac{\partial_x \Psi-2\partial_x \psi }{\sqrt{\kappa_2^2-\tau_2^2}} \,. 
\end{align}
The constancy of $\tau_2$ yields the Lorentzian Liouville equation \eqref{null_Liouvilleeq}
for $\psi$ with $k=-2 \tau_2^2$. 
Defining $x=\frac{\sqrt 2}{\tau_2}e^{\phi} \tanh (r/2)$ and 
$y=\frac{\sqrt 2}{\tau_2}e^{-\phi} \tanh (r/2)$, the orthonormal frame simplifies to 
\begin{align}
\label{[1,(11)]metric:tau2+tau3=0:tau2sq-kappa2sq<0}
e_1 =&\, -(\D t-t_0 \cosh r\D \phi ) \,, \qquad 
e_3=\frac{t_0}{\sqrt 2 }\left[\sinh(t/t_0)\D r + \cosh (t/t_0)\sinh r\D \phi \right]\,, \notag \\
e_2=&\, \frac{t_0 }{\sqrt 2}{\rm csch}\beta  \left[
\cosh(t/t_0)(\D r -\cosh \beta\sinh r \D \phi)
-\sinh(t/t_0)(\cosh\beta \D r-\sinh r\D \phi )
\right] \,, 
\end{align}
where we have shifted the time coordinates by 
$t\to t-t_0 [\Psi+\phi-2 \log (\cosh(\frac{r}{2}))]$ 
and defined new constants ($t_0, \beta$) by
$\kappa_2=t_0^{-1}\cosh\beta$, $\tau_2 =t_0^{-1} \sinh\beta $. 
The 3 KVs are given by 
\begin{subequations}
\begin{align}
\label{}
K_1 =& \,- t_0 \cosh \phi\,{\rm csch }\, r\partial_t +\sinh\phi \partial_r -\cosh\phi \coth r\partial_\phi \,, \\ 
K_2=& \, t_0 \sinh\phi \,{\rm csch}\, r\partial_t -\cosh\phi \partial_r +\sinh\phi\, \coth r\partial_\phi \,,  \\ 
K_3=&\,  \partial_\phi \,, 
\end{align}
\end{subequations}
satisfying $\mathfrak{so}(1,2)$ algebra
\begin{align}
\label{}
[K_1, K_2]=- K_3 \,, \qquad
[K_2, K_3]=K_1 \,, \qquad 
[K_3, K_1]=-K_2 \,, 
\end{align}

\subsubsection{$\tau_2^2-\kappa_2^2>0$ and $\tau_2\ne 0$}

The equation (\ref{[1,(11)]_alphaeq}) admits two distinct roots 
$\alpha= \tau_2^{-1}(\kappa_2 + i  \sqrt{\tau_2^2-\kappa_2^2}) $ and its 
complex conjugation, so that 
$W_a=e_{2a}+\alpha e_{3a}$ satisfies
\begin{align}
\label{}
 \nabla_b W_a =  i\sqrt{\tau_2^2 -\kappa_2^2}e_{1a} W_b \,, \qquad 
 W_a \bar W^a= 2 \,. 
\end{align}
One can thus introduce coordinates 
\begin{align}
\label{}
e_{1a}=-(\nabla_a t+\chi_1 \nabla_a  x+\chi_2 \nabla_a y )\,, \qquad 
W_a =e^{\psi+i \theta }(\nabla_a x+i \nabla_b y ) \,. 
\end{align}
where  $\chi_1, \chi_2, \psi, \theta $ are functions of $t, x,y$.

From $\kappa_2+\kappa_3=0 $ and $\kappa_2={\rm constant}$, 
one finds $\psi=\psi(x,y)$ and $\theta=t \sqrt{\tau_2^2-\kappa^2_2} +\Psi(x,y)$. 
The conditions $\kappa_1=\eta_1=0$ give 
$\partial_t \chi_1=\partial_t \chi_2=0$. Plugging this into $\eta_3=0$, 
one finds
\begin{align}
\label{}
\chi_1=\frac{\partial_y \psi+\partial_x \Psi}{\sqrt{\tau_2^2-\kappa_2^2}} \,, \qquad 
\chi_2=\frac{-\partial_x \psi+\partial_y \Psi}{\sqrt{\tau_2^2-\kappa_2^2}} \,. 
\end{align}
$\tau_2={\rm const.}$ requires 
that $\psi=\psi(x,y)$ obeys Liouville's equation (\ref{Liouvilleeq}) 
with $k \equiv -2 \tau_2^2 <0$. 
The solution to Liouville's equation can be chosen to be (\ref{Liouvillesol}). 
By the change of the coordinates 
$x+i y=\frac {\sqrt 2}{\tau_2}\tanh(\frac{r}{2}) e^{i\phi}$ with 
$t\to t-(\Psi+\phi) /\sqrt{\tau_2^2-\kappa_2^2}$, 
one can bring the triad frame to 
\begin{align}
\label{[1,(11)]metric:tau2+tau3=0:tau2sq-kappa2sq>0}
e_1= & -(\D t-t_0 \cosh r \D \phi) \,, \qquad 
e_3= \frac{t_0}{\sqrt 2}\left(\sin(t/t_0 ) \D r+\cos (t/t_0) \sinh r \D \phi \right) \,, \notag \\
e_2=& \frac{t_0}{\sqrt 2} {\rm sech} \beta \left[\cos(t/t_0) \left(\D r- \sinh \beta \sinh r  \D \phi \right)
-\sin(t/t_0) \left(\sinh \beta \D r+\sinh r \D \phi \right)
 \right]\,, 
\end{align}
where we have redefined constants as 
$\tau_2=t_0^{-1} \cosh\beta $, $\kappa_2= t_0^{-1} \sinh\beta$. 
The KVs are 
\begin{align}
\label{}
K_1=&\,  t_0 \cos\phi \,{\rm csch}r \partial_t +\sin\phi  \partial_r +\cos\phi \,{\rm coth}r \partial_\phi \,, \notag \\ 
K_2=& \, t_0 \sin \phi \,{\rm csch}r \partial_t -\cos \phi  \partial_r +\sin\phi \,{\rm coth}r \partial_\phi \,, \\ 
K_3= &\,  \partial_ \phi \,. \notag
\end{align}
These KVs form the $\mathfrak{so}(1, 2)$ algebra
\begin{align}
\label{}
[K_1, K_2]=K_3 \,, \qquad 
[K_2, K_3]=-K_1 \,, \qquad 
[K_3, K_1]=-K_2 \,.
\end{align}

\subsection{$\tau_2+\tau_3\ne0 $}
\label{sec_[1,(11)]_case2}

In this branch, the second obstruction matrix is given by (4.13c) in \cite{Nozawa:2019dwu}. 
This vanishes provided 
\begin{align}
\label{[1,(11)]_2ndobs}
\tau_- ={\rm constant}\,, \qquad \Pi_i^{[1,(11)]}=\Xi_i^{[1,(11)]}=0 \,, 
\end{align}
where 
\begin{subequations}
\begin{align}
\label{}
\tau_\pm \equiv\, &\tau_3 \pm \tau_2 \,, \\ 
\Pi_i^{[1,(11)]} \equiv \,&\ma L_i \eta_2 +\frac{\ma L_i \ma L_2 \kappa_2}{\tau_+} - \left(
\ma L_2(\tau_-+3\tau_+)-4 \eta_2 \kappa_2
\right)\frac{\ma L_i \kappa_2}{2\tau_+^2}  \,, \\
\Xi_i^{[1,(11)]} \equiv\, & \ma L_i \eta_3 {-\frac{\ma L_i \ma L_3 \kappa_2}{\tau_+}} - \left(
\ma L_3(\tau_--3\tau_+)-4 \eta_3 \kappa_2
\right)\frac{\ma L_i \kappa_2}{2\tau_+^2}  \,. 
\end{align}
\end{subequations}

Plugging $\tau_- ={\rm const.}$ into the 
Ricci identity (\ref{eq:app_ricci_ortho}) and using the Jordan normal form (\ref{Segre[1,(11)]}), 
we obtain the 1st-order system for $\tau_2$: 
\begin{align}
\label{[1,(11)]_tau2eq}
\ma L_1\tau_2 =2\kappa_2 \tau_1 \,, \qquad 
\ma L_2 \tau_2 =2 \eta_2 \kappa_2 +\eta_3(\tau_2+\tau_3)-\ma L_3 \kappa_2 \,, \qquad 
\ma L_3 \tau_2 =-2\eta_3 \kappa_2+\eta_2 (\tau_2+\tau_3)+\ma L_2 \kappa_2 \,. 
\end{align}
The compatibility \eqref{eq:app_comm_ortho} of these equations gives rise to 
\begin{align}
\label{[1,(11)]_kappa2eq}
\ma L_2 \kappa_2 =-\eta_2 (\tau_2+\tau_3)\,, \qquad \ma L_3 \kappa_2 =\eta_3(\tau_2+\tau_3)\,. 
\end{align}
Further compatibility of these equations yields 
\begin{align}
\label{[1,(11)]_lambda}
\lambda =\frac 23(\kappa_2^2+\tau_2\tau_3) \ne 0\,. 
\end{align}
One can verify that $\Pi_i^{[1,(11)]} =\Xi_i^{[1,(11)]} =0$ are all fulfilled.

Thanks to the type II gauge transformations (\ref{nonnull_localSO21_II}), 
one can always achieve $\kappa_2=0$, while  eqs. (\ref{[1,(11)]_2ndobs}), (\ref{[1,(11)]_tau2eq}), (\ref{[1,(11)]_kappa2eq}) and 
(\ref{[1,(11)]_lambda}) remain unchanged. In this gauge, we have 
$\eta_2=\eta_3=\tau_1=0$ and $\tau_2={\rm const}$. 
Then, the orthonormal frame obeys
\begin{align}
\label{}
\nabla_b e_{1a}= \tau_2 e_{3a}e_{2b}+ \tau_3 e_{2a}e_{3b} \,, \qquad 
\nabla_b e_{2a}= \tau_3 e_{1a} e_{3b} \,, \qquad 
\nabla_b e_{3a}= \tau_2 e_{1a} e_{2b} \,. 
\end{align}
The local metrics fall into two further subclasses depending on $\tau_2\tau_3 \gtrless 0$. 
The $\tau_2\tau_3 =0$ possibility is excluded by (\ref{[1,(11)]_lambda}). 
Since the solution possesses two  constant Ricci rotation coefficients  ($\tau_2, \tau_3$), 
the metric is homogeneous.

\subsubsection{$\tau_2\tau_3 > 0$}
\label{sec_[1,(11)]_case2_tau23>0}

In this case, two real vectors
$W_a^\pm =e_{2a}\pm  \sqrt{\tau_3 /\tau_2}e_{3a}$ are hypersurface-orthogonal $W_{[a}^\pm\nabla_b W^\pm_{c]}=0$, 
which allows us to find local coordinates ($t, x, y$) such that 
\begin{align}
\label{}
e_{1a}=- (\nabla_a t +\chi_1 \nabla_a x+\chi_2\nabla_a y)\,, \qquad 
W^+_a =  e^{\phi_+ }\D x \,, \qquad 
W^-_a=  e^{\phi_- }\D y\,.
\end{align}
Substituting these into $\kappa_2=\kappa_3=0$, 
we have 
$\phi_-(t,x,y)=-\phi_+(t,x,y)+2\psi (x,y)$. 
Conditions $\kappa_1=\eta_1=0$ now yield $\partial_t \chi_1=\partial_t \chi_2=0$, whereas
the constancy of $\tau_2, \tau_3$ gives
\begin{align}
\label{}
\partial_y \chi_1 = \partial_x \chi_2 +\frac 12 (\tau_3-\tau_2)\left(\frac{\tau_3}{\tau_2}\right)^{-1/2} e^{2\psi} \,, \qquad 
\phi_+(t,x,y)= \tau_2 \sqrt{\frac{\tau_3}{\tau_2}} t+\Psi(x,y) \,. 
\end{align}
Inserting these into $\eta_2=\eta_3=0$, we find
\begin{align}
\label{}
\partial_x \Psi= \tau_2\sqrt{\frac{\tau_3}{\tau_2}} \chi_1+2\partial_x \psi \,, \qquad 
\partial_y \Psi =\tau_2\sqrt{\frac{\tau_3}{\tau_2}} \chi_2 \,. 
\end{align}
The integrability $(\partial_x\partial_y -\partial_y \partial_x)\Psi=0$ yields Lorentzian 
Liouville's equation (\ref{null_Liouvilleeq}) for $\psi$ with $k=\tau_2(\tau_3-\tau_2)$, whence  
we can choose $\psi $ as (\ref{null_Liouvillesol}). 
To simplify the form of the metric, let us define
\begin{align}
\label{}
k \equiv  \tau_2(\tau_3-\tau_2 )\,, \qquad 
t_0^{-1}\equiv  \tau_2 \left(\frac{\tau_3}{\tau_2}\right)^{1/2}\,, \qquad 
x=\frac{2}{\sqrt k}\tan \left(\frac{\sqrt k}{2} r \right)e^{\phi} \,, \qquad 
y=\frac{2}{\sqrt k}\tan \left(\frac{\sqrt k}{2} r \right)e^{-\phi} \,,
\end{align}
and shift the time coordinate as 
$t\to t-t_0 [\Psi+\phi-2 \log (\cos (\frac{\sqrt k}{2}r))]$. 
Then the triad frame reduces to 
\begin{subequations}
\label{[1,(11)]metric:tau_2+tau3ne0:tau2tau3>0}
\begin{align}
e_{1} =&\, - \left(\D t -t_0 \cos(\sqrt kr) \D \phi \right) \,, \\ 
e_2 =&\, \cosh(t/t_0)\D r+\frac{\sin (\sqrt k r)}{\sqrt k}\sinh (t/t_0)\D \phi \,,\\
e_3=&\, \left(\frac{\tau_3}{\tau_2}\right)^{-1/2} \left(\sinh(t/t_0)\D r+\frac{\sin (\sqrt k r)}{\sqrt k}\cosh (t/t_0)\D \phi \right)\,. 
\end{align}
\end{subequations} 
These expressions are well-defined irrespective of the sign of $k$. 
The KVs are given by 
\begin{subequations}
\begin{align}
\label{}
K_1 =&\,
-\sqrt kt_0 \cosh \phi \,{\rm csc}(\sqrt k r)\partial_t +\sinh \phi \partial_r -\sqrt k \cosh \phi \cot (\sqrt k r)\partial_\phi \,, \\ 
K_2=&\, \sqrt kt_0 \sinh \phi\, {\rm csc}(\sqrt k r)\partial_t -\cosh \phi \partial_r +\sqrt k \sinh \phi \cot (\sqrt k r)\partial_\phi \,, \\ 
K_3=&\, \partial_\phi \,, 
\end{align}
\end{subequations}
satisfying 
\begin{align}
\label{}
[K_1, K_2]= k K_3 \,, \qquad [K_2, K_3]=  K_1\,, \qquad 
[K_3, K_1]=-K_2 \,. 
\end{align}

\subsubsection{$\tau_2\tau_3< 0$}
\label{sec_[1,(11)]_case2_tau23<0}

Since the complex vector $W_a =e_{2a}+i \sqrt{-\tau_3/\tau_2}e_{3a}$ is hypersurface-orthogonal, 
one can introduce the coordinates as 
\begin{align}
\label{}
e_1=-(\D t+\chi_1 \D x+\chi_2 \D y )\,, \qquad 
W=e^{\psi +i \theta} (\D x+i \D y )\,. 
\end{align}
$\psi $ and $\chi_{1,2}$ are $t$-independent on account of 
$\kappa_1=\eta_1=\kappa_2=0$. The constancy of $\tau_{2,3}$ 
amounts to $\theta=t \tau_2\sqrt{-\tau_3/\tau_2} +\Psi(x,y)$ and 
$\partial_y \chi_1=\partial_x \chi_2 +(\tau_2-\tau_3)(-\tau_3/\tau_2)^{-1/2}e^{2\psi} $. 
One finds a set of equation for $\Psi$ from $\eta_2=\eta_3=0$, whose integrability 
corresponds to the Liouville equation for $\psi$ (\ref{Liouvilleeq}) with 
$k=\tau_2(\tau_3 -\tau_2)$. 
Defining 
\begin{align}
\label{}
k\equiv \tau_2(\tau_3-\tau_2) \,, \qquad t_0^{-1} \equiv \tau_2 \sqrt{-\frac{\tau_3}{\tau_2}}\,, 
 \qquad x+i y =\frac{2}{\sqrt k}\tan \left(\frac{\sqrt k}{2}r\right) e^{i\phi} \,, 
\end{align}
and changing the time coordinate by $t\to t- t_0(\Psi+\phi)$, 
the orthonormal frame is tantamount to 
\begin{subequations}
\label{[1,(11)]metric:tau_2+tau3ne0:tau2tau3<0}
\begin{align}
e_1= & \, - \left(\D t -t_0 \cos (\sqrt k r)\D \phi \right)\,, \\
e_2=&\,  \cos(t/t_0)\D r-\frac{\sin(\sqrt k r)}{\sqrt k}\sin(t/t_0)\D \phi \,, \\ 
e_3=&\, \left(-\frac{\tau_3}{\tau_2}\right)^{-1/2} \left(\sin(t/t_0)\D r+\frac{\sin(\sqrt k r)}{\sqrt k}\cos(t/t_0)\D \phi \right)\,. 
\end{align}
\end{subequations} 
The 3 KVs are given by 
\begin{subequations}
\begin{align}
\label{}
K_1=& \, \sqrt kt_0 \cos\phi \,{\rm csc}(\sqrt kr)\partial_t +\sin\phi \partial_r +\sqrt k\cos\phi \cot(\sqrt kr)\partial_\phi \,, \\
K_2=& \, -\sqrt k t_0  \sin\phi \,{\rm csc}(\sqrt kr)\partial_t +\cos\phi \partial_r -\sqrt k\sin\phi \cot(\sqrt kr)\partial_\phi \,, \\
K_3=& \, \partial_\phi \,, 
\end{align}
\end{subequations} 
satisfying 
\begin{align}
\label{}
[K_1, K_2]= k K_3 \,, \qquad 
[K_2, K_3]= K_1 \,, \qquad 
[K_3 , K_1]=K_2 \,. 
\end{align}

\section{Homogeneous metrics: Segre [(1,1)1]} 
\label{sec_[(1,1)1]}

This section considers the metrics belonging to the Segre [(1,1)1]. 
As deduced from the results in section \ref{subsec:inhom1}, 
this case corresponds to the double Wick rotation of the [1,(11)] case. 
Therefore, we shall not discuss the derivation in detail, but outline the argument and show quick results. 

The trace-free part of the Ricci tensor takes the form
\begin{align}
\label{}
S_{ab}=-\lambda (e_{1a}e_{1b} +2e_{2a}e_{2b}-e_{3a}e_{3b})\,, 
\end{align}
where the metric is given by (\ref{nonnull_frame}) with $\varepsilon=-1$. 
The Bianchi identity implies 
\begin{align}
\label{}
\kappa_2=0 \,, \qquad \eta_2=0 \,, \qquad \eta_3 =\kappa_1\,, 
\end{align}

\subsection{$\tau_1=\tau_3$}
\label{sec_[(1,1)1]_case1}

When
\begin{align}
\label{}
\tau_1=\tau_3\,, \qquad \kappa_1 \ne 0 \,, 
\end{align}
the second obstruction matrix is given by (4.19c) of \cite{Nozawa:2019dwu}, 
which vanishes iff 
\begin{align}
\label{}
\{\tau_1, \eta_1 , \kappa_3\}={\rm constants} \,. 
\end{align}
From the constancy of $R_{22}$, $\kappa_1$ turns out to be a nonvanishing constant.
The Jordan form of $S_{ab}$ then requests $\kappa_3=\eta_1=\tau_2=0$. 
The curvature tensors reduce to 
\begin{align}
\label{}
\lambda=\frac 23 (\kappa_1^2+\tau_1^2 ) \,, \qquad R=-2(\kappa_1^2+\tau_1^2)\,. 
\end{align}

\subsubsection{$\tau_1=0$}

When $\tau_1=0$, $e_1$ and $e_3$ are hypersurface-orthogonal and 
$e_2$ is closed, 
for which the final metric reads
\begin{align}
\label{[(1,1)1]metric:tau1=tau3:tau1=0}
\D s^2=-e^{2\kappa_1 x}\D t^2+ \D x^2+ e^{-2\kappa_1 x}\D y^2 \,. 
\end{align}
The KVs are given by
\begin{align}
\label{}
K_1=\partial_t \,, \qquad K_2=\partial_y \,, \qquad 
K_3=t \partial_t -\kappa_1^{-1}\partial_x-y\partial_y \,, 
\end{align}
with the commutators 
\begin{align}
\label{}
[K_2, K_3]=-K_2\,, \qquad [K_3, K_1]=- K_1 \,. 
\end{align}

\subsubsection{$\tau_1\ne 0$}

In this case, the vectors $W^\pm_{a} = e_{1a}+\tau_1^{-1}(\kappa_1\pm \sqrt{\kappa_1^2+\tau_1^2})e_{3a}$
are hypersurface-orthogonal, and the final orthonormal frame boils down to 
\begin{subequations}
\label{[(1,1)1]metric:tau1=tau3:tau1ne0}
\begin{align}
e_1=&\, \frac{x_0}{\sqrt 2}\csc\beta \left[(\D t-\cos\beta \sin t \D \phi) \cosh (x/x_0)+
(\cos\beta \D t-\sin t \D \phi )\sinh(x/x_0) \right]\,, \\
e_{2}=&\,\D x+x_0\cos t \D \phi \,, \qquad 
e_3=\frac {x_0}{\sqrt 2}\left(-\sinh(x/x_0)\D t+\cosh(x/x_0)\sin t\D \phi \right)\,, 
\end{align}
\end{subequations}
where we have defined 
$\kappa_1=x_0^{-1} \cos\beta $, $\tau_1=x_0^{-1}\sin \beta $. 
The KVs are given by
\begin{subequations}
\begin{align}
\label{}
K_1=&\, \sinh \phi\partial_t +x_0\cosh\phi \csc t \partial_x-\cosh\phi \cot t \partial_\phi \,, \\
K_2=&\,  -\cosh\phi \partial_t -x_0\sinh\phi\csc t\partial_x+\sinh\phi \cot t\partial_\phi\,, \\
K_3=&\,  \partial_\phi \,, 
\end{align}
\end{subequations}
satisfying the $\mathfrak{so}(1,2)$ algebra 
\begin{align}
\label{}
[K_1, K_2]=K_3 \,, \qquad [K_2, K_3]=K_1\,, \qquad 
[K_3, K_1]=-K_2 \,. 
\end{align}

\subsection{$\tau_1\ne \tau_3$}
\label{sec_[(1,1)1]_case2}

The second obstruction matrix in this case reads  (4.20c) in \cite{Nozawa:2019dwu}, 
whose vanishing requires 
\begin{align}
\label{}
\tau_+={\rm constant} \,, \qquad \Pi_i^{[(1,1)1]}=\Xi_i^{[(1,1)1]}=0\,, 
\end{align}
where  $\tau_\pm =\tau_3\pm \tau_1$ and 
\begin{align}
\label{}
\Pi_i^{[(1,1)1]} \equiv \, & \ma L_i \eta_1-\frac{\ma L_i \ma L_1 \kappa_1}{\tau_-}+\Bigl(
\ma L_1 (\tau_++3\tau_-)-4\eta_1 \kappa_1\Bigr)\frac{\ma L_i\kappa_1}{2\tau_-^2} \,, \\
\Xi_i^{[(1,1)1]}  \equiv \, & \ma L_i \kappa_3+\frac{\ma L_i\ma L_3 \kappa_1}{\tau_-}+ 
\Bigl(\ma L_3 (\tau_+-3\tau_-)-4\kappa_1\kappa_3\Bigr)\frac{\ma L_i \kappa_1}{2\tau_-^2} \,. 
\end{align}

Inserting of $\tau_+={\rm constant}$ into the Ricci identity, 
one gets
\begin{align}
\label{}
\ma L_1 \tau_1=-2\eta_1 \kappa_1 +\kappa_3(\tau_1-\tau_3)-\ma L_3 \kappa_1 \,, \qquad 
\ma L_2 \tau_1=-2\kappa_1 \tau_2\,, \qquad 
\ma L_3 \tau_1=2\kappa_1\kappa_3 +\eta_1(\tau_3-\tau_1)-\ma L_1 \kappa_1 \,. 
\end{align}
The compatibility conditions (\ref{eq:app_comm_ortho}) for  these equations yield
\begin{subequations}
\label{[(1,1)1]_compeq}
\begin{align}
(\tau_1-\tau_3)[\ma L_1 \kappa_1-2\kappa_1 \kappa_3+\eta_1(\tau_1-\tau_3)]+2\kappa_1 \ma L_3\kappa_1 =&0 \,, \\
2\kappa_1 \ma L_1 \kappa_1+(\tau_1-\tau_3)[\ma L_3 \kappa_1 +2\eta_1\kappa_1-\kappa_3(\tau_1-\tau_3)]=&0\,. 
\end{align}
\end{subequations}
The following analysis proceeds according to $4\kappa_1^2=(\tau_1-\tau_3)^2$ or not.

\subsubsection{$4\kappa_1^2\ne (\tau_1-\tau_3)^2$}

If $4\kappa_1^2\ne (\tau_1-\tau_3)^2$, equations
(\ref{[(1,1)1]_compeq}) can be solved to give $\ma L_1 \kappa_1=\eta_1(\tau_3-\tau_1)$ 
and $\ma L_3 \kappa_1=\kappa_3(\tau_1-\tau_3)$, 
whose additional compatibility condition yields
$\lambda =\frac 23(\kappa_1^2+\tau_1\tau_3)$. 
This implies that $\Pi_i^{[(1,1)1]}=\Xi_i^{[(1,1)1]}=0$ are satisfied trivially.

If $|2\kappa_1/(\tau_1-\tau_3)|>1$, one can use the gauge freedom (\ref{nonnull_localSO21_III})
to bring about $\tau_1=\tau_3$. Since this criterion has been already discussed in the previous subsection \ref{sec_[(1,1)1]_case1}, 
we do not discuss this case here.  
Supposed $|2\kappa_1/(\tau_1-\tau_3)|<1$, 
one can always achieve $\kappa_1=0$ using the gauge freedom (\ref{nonnull_localSO21_III}). This leads to 
$\eta_1=\kappa_3=\tau_2=0$ and constancy of $\tau_1$.
The nonvanishing Ricci rotation coefficients are ($\tau_1, \tau_3$), which 
are constants satisfying $\tau_1 (\tau_3-\tau_1)\ne 0$. Repeating the argument in section \ref{sec_[1,(11)]_case2}, 
the final orthonormal frame reduces to 
\begin{align}
\label{[(1,1)1]metric:tau1netau3:4kappa1sqne(tau1-tau3)sq}
e_1=&\cosh (\sqrt{\epsilon}x/x_0)\D t-\frac {\sqrt{\epsilon}}{\sqrt k}\sin(\sqrt k t)\sinh(\sqrt{\epsilon}x/x_0)\D \phi \,, \qquad 
e_2=\D x+x_0 \cos(\sqrt k t)\D \phi \,, \notag \\
e_3=& \sqrt{\epsilon}\sqrt{\frac{\tau_1}{\tau_3}}\left(-\frac 1{\sqrt{\epsilon}}\sinh(\sqrt{\epsilon}x/x_0)\D t+\frac 1{\sqrt k}\sin(\sqrt k t)\cosh(\sqrt{\epsilon}x/x_0)\D \phi 
\right)\,, 
\end{align}
 where we have denoted
 \begin{align}
\label{}
k\equiv \tau_1(\tau_1+\tau_3) \,, \qquad x_0=\frac{\sqrt{\epsilon}}{\tau_3}\left(\frac{\tau_1}{\tau_3}\right)^{-1/2}\,, \qquad 
\epsilon={\rm sgn}(\tau_1/\tau_3) =\pm 1\,. 
\end{align}
The KVs are given by 
\begin{subequations}
\begin{align}
\label{}
K_1=& \, \sqrt \epsilon \sinh(\sqrt \epsilon \phi)\partial_t +x_0 \sqrt k \cosh(\sqrt \epsilon\phi)\csc (\sqrt k t)\partial_x
-\sqrt k\cosh (\sqrt \epsilon\phi)\cot (\sqrt k t) \partial_\phi \,, \\
K_2=& \, -\cosh(\sqrt \epsilon\phi)\partial_t -x_0 \frac{\sqrt k}{\sqrt \epsilon} \csc (\sqrt kt)\sinh(\sqrt \epsilon\phi)\partial_x +\frac{\sqrt k}{\sqrt \epsilon}
\cot (\sqrt k t)\sinh (\sqrt \epsilon\phi)\partial_\phi \,, \\
K_3=& \, \partial_\phi \,, 
\end{align}
\end{subequations}
satisfying
\begin{align}
\label{}
[K_1, K_2]=k K_3 \,, \qquad [K_2, K_3]=K_1 \,, \qquad [K_3,K_1]=-\epsilon K_2 \,. 
\end{align}

\subsubsection{$4\kappa_1^2=(\tau_1-\tau_3)^2$}

When $4\kappa_1^2=(\tau_1-\tau_3)^2$, equations in (\ref{[(1,1)1]_compeq}) are 
not independent. The gauge transformation (\ref{nonnull_localSO21_III}) enables us to 
employ the frame for which $\kappa_1$ is a nonvanishing constant. 
This leads to the constancy of $\tau_1$ and $\tau_3=\tau_1\pm 2 \kappa_1$. From the Ricci identity in the Jordan form, 
we have $\tau_2=0$ and $\kappa_3=\mp \eta_1$. Combining the Ricci identity with  $\Pi_i^{[(1,1)1]}=0$,
we get $\eta_1=0$. Thus, the solution is characterized by 2 parameters ($\kappa_1, \tau_1$). 
The curvature tensors are reduced to 
\begin{align}
\label{}
\lambda =\frac 23 (\kappa_1\pm \tau_1)^2 \,, \qquad R=-2(\kappa_1\pm \tau_1)^2 \,. 
\end{align}

The metric is written as \cite{Nozawa:2019dwu} 
\begin{subequations}
\label{[(1,1)1]metric:tau1netau3:4kappa1sq=(tau1-tau3)sq}
\begin{align}
e_1=& \cosh(x/x_0)\D t-\frac{\sin(\sqrt kt)}{\sqrt k}\sinh(x/x_0)\D \phi
+x_0 \kappa_1 \Bigl(\sinh(x/x_0)\D t-\frac{\sin(\sqrt kt)}{\sqrt k}\cosh(x/x_0)\D \phi \Bigr)\,, \\
e_2=&\D x+x_0 \cos (\sqrt k t) \D \phi \,, \\
e_3=& x_0 \left(-\tau_1 \sinh(x/x_0)\D t\pm \frac 12 x_0 \sqrt k \sin(\sqrt k t)\cosh(x/x_0)\D \phi\right)\,,
\end{align}
\end{subequations}
where we have denoted 
\begin{align}
\label{}
k\equiv  2\tau_1(\tau_1\pm \kappa_1)\,, \qquad x_0^{-1}\equiv \kappa_1\pm \tau_1\,. 
\end{align}
The KVs are given by
\begin{subequations}
\begin{align}
\label{}
K_1=&\, \sinh \phi \partial_t +\frac{2\tau_1}{\sqrt k}\csc(\sqrt k t)\cosh \phi \partial_x -\sqrt k \cot (\sqrt kt)\cosh\phi
\partial_\phi \,, \\
K_2=&\, -\cosh \phi \partial_t \mp \frac{2\tau_1}{\sqrt k}\csc(\sqrt k t)\sinh \phi \partial_x +\sqrt k \cot (\sqrt kt)\sinh\phi
\partial_\phi \,, \\
K_3=& \, \partial_\phi \,, 
\end{align}
\end{subequations}
with the commutators
\begin{align}
\label{}
[K_1, K_2]=k K_3 \,, \qquad 
[K_2, K_3]=K_1 \,, \qquad [K_3, K_1]=-K_2 \,. 
\end{align}

\section{Homogeneous metrics: Segre [1,11]} 
\label{sec:[1,11]}

In this class, 
the trace-free Ricci tensor takes the form 
\begin{align}
\label{[1,11]_Sab}
S_{ab}=-\lambda_1 e_{1a}e_{1b}+ \lambda_2 e_{2a}e_{2b}+\lambda_3 e_{3a}e_{3b} \,, \qquad 
\lambda_1+\lambda_2+\lambda_3=0 \,, 
\end{align}
where all eigenvalues $\lambda_i$ are constant and take distinct values. From the above form of $S_{ab}$, 
the only remaining degrees of freedom for the frame choice is the exchange of $e_2 \leftrightarrow e_3$ with 
$\lambda _2 \leftrightarrow  \lambda_3$ corresponding to (\ref{nonnull_localSO21_II}) with $a_2=\pi/2$.
The second obstruction matrix is given by (4.24) of \cite{Nozawa:2019dwu}, 
which vanishes iff 
\begin{align}
\label{}
\{\kappa_1, \kappa_2, \eta_1 ,\tau_1, \tau_2 , \tau_3\}={\rm constants.} 
\end{align}
The Bianchi identity puts further restrictions on Ricci rotation coefficients 
\begin{align}
\label{[1,11]_Bianchi}
\kappa_3= \frac{\lambda_1-\lambda_2}{\lambda_3-\lambda_1} \kappa_2 \,, \qquad 
\kappa_1= -\frac{\lambda_2-\lambda_3}{\lambda_1-\lambda_2} \eta_3 \,, \qquad 
\eta_2 =-\frac{\lambda_3-\lambda_1}{\lambda_2-\lambda_3} \eta_1 \,. 
\end{align}
It then follows that all the Ricci rotation coefficients are constants, leading to the 
homogeneity of the solution. 

The condition $R_{ab}=R_{ba}$ 
culminates in a zero eigenvalue equation 
\begin{align}
\label{}
 N^{ab} X_b=0 
\end{align}
where the symmetric matrix $N^{ab}=N^{ba}$ and the vector 
$X_b$ are given by 
\begin{align}
\label{NabXb}
N^{ab} \equiv\left(
 \begin{array}{ccc}
2( \tau_2-\tau_3)  & -\eta_1-\eta_2 & \kappa_1 +\eta_3     \\
-\eta_1-\eta_2  & 2(\tau_1+\tau_3) & \kappa_2-\kappa_3  \\
\kappa_1 +\eta_3  & \kappa_2 -\kappa_3 & 2(\tau_1-\tau_2 )
\end{array}
 \right)\,, \qquad 
 X_a= \left(
 \begin{array}{c}
\kappa_2+\kappa_3  \\
\eta_3-\kappa_1 \\ 
\eta_2-\eta_1
\end{array}
 \right) \,. 
\end{align}
Let $C^k{}_{ij}=C^k{}_{[ij]}$ be a collection of Ricci rotation coefficients such that 
$[e_i, e_j]=C^k{}_{ij} e_k$, c.f., \eqref{eq:app_comm_ortho}. Since $C^k{}_{ij}$ are all constants, one can view 
$C^k{}_{ij}$ as the structure constants of the algebra and $e_i$ as invariant dual forms. 
Then $N^{ab}X_b=0$ is nothing but the condition that 
the matrices $(\bs C_i)^k{}_j =C^k{}_{ij}$ constitute the adjoint representation of the algebra
$[\bs C_i, \bs C_j]=-C^k{}_{ij}\bs C_k$. 
This is the Lorentzian version of the classification of Bianchi cosmology~\cite{Stephani:2003tm,Bianchi:1898,Ellis:1968vb}.

The standard method for the classification of Bianchi type is the diagonalization of $N^a{}_bX^b=0$. However, 
we do not attempt to follow this route by the 2 reasons. The first rationale is that the diagonalization of $N^a{}_b$ 
is a cumbersome task since $N^a{}_b$ is no longer a symmetric matrix in the Lorentzian signature. Secondly, we wish to 
keep  the Jordan form (\ref{[1,11]_Sab}) of the traceless Ricci tensor. Further  diagonalization of $N^a{}_b$ does not preserve this property. 
In lieu of diagonalization, we shall investigate all possible cases of zero eigenvalue problem (\ref{NabXb}) by utilizing (\ref{[1,11]_Bianchi}), and check whether $S_{ab}$ is of diagonal form. 

Let us first separate our inquiry  into two categories  (A) $\kappa_2 =0$ or (B) $\kappa_2\ne 0$. 
In the case (A), the first component of (\ref{NabXb}) reduces to $\eta_1\eta_3 \lambda_1=0$, which falls into 
(i) $\eta_1=0$, (ii) $\eta_3=0$ with $\eta_1\ne 0 $, (iii) $\lambda_1=0$ with $\eta_1\eta_3\ne 0 $. In the case (A-i), the second  component of (\ref{NabXb})
requires $\eta_3 (\tau_1+\tau_3)=0$, which is further distinguished into (1) $\eta_3=0$ and (2) $\tau_1+\tau_3=0$ with $\eta_3\ne 0$. 
$S_{ab}$ becomes diagonal for (A-i-1), whereas for (A-i-2) $S_{13}=0$ demands $\tau_1=-3\lambda_2\tau_2/(\lambda_1-\lambda_3)$. 
For (A-ii), we need $\tau_2=\tau_1$ from (\ref{NabXb}) and $S_{12}=0$ asks for $\tau_1=3\lambda_3\tau_3/(\lambda_1-\lambda_2)$. 
For (A-iii), we have $\tau_1=\tau_2=\tau_3=0$  from (\ref{NabXb})  and $S_{13}=0$ demands $\eta_1\eta_3=0$, which is a contradiction to the assumption $\eta_1\eta_3\ne 0 $. 
Hence the class (A-iii) does not admit solutions with 3 KVs.

For case (B) $\kappa_2\ne 0$, the first  component of (\ref{NabXb})  can be solved with respect to $\tau_3$. 
We split the analysis according to (i) $\eta_3=0$ and (ii) $\eta_3 \ne 0$. 
For (i) $\eta_3=0$, we have $\tau_3=\tau_2$ and  $\eta_1(\tau_1-\tau_2)=0$ by the third component of (\ref{NabXb}), leading to the subclasses (1) $\eta_1=0$ and (2) $\tau_2=\tau_1$ with $\eta_1\ne 0$. 
For (B-i-1) $\tau_2=-3\lambda _1 \tau_1/(\lambda_2-\lambda_3) $ follows from $S_{23}=0$.  
For (B-i-2), we obtain $\lambda_2=0$ from the second  component of (\ref{NabXb}). Then we have $S_{13}=-3\eta_1 \kappa_2 \ne 0$, 
inconsistent with the Jordan form. So the class (B-i-2) has no solutions. 
Consider next (B-ii), for which $\kappa_2\eta_3 \ne 0$. Then, the second component of (\ref{NabXb}) is solved to give $\tau_2$. Supposed $\eta_1=0$, $S_{12}=0$ is never satisfied, so that we have $\eta_1\ne 0$. Then the third component of  (\ref{NabXb}) is solved to give $\tau_1$. After some computations, one finds that the rest of the condition for the Jordan form (\ref{[1,11]_Sab}) fails to be fulfilled. Hence, we have no solutions in the class (B-ii). 

We thus obtained four classes to consider (A-i-1), (A-i-2), (A-ii) and (B-i-1). 
However, we have not yet made use of the freedom 
for the discrete frame change $e_2 \leftrightarrow e_3$. As it turns out, 
the (A-i-2) case is equivalent to (A-ii) under this frame change. 
We shall therefore consider separately  
(A-i-1),  (A-ii) and (B-i-1) in the following. 

\subsection{case (A-i-1): $\kappa_2=\eta_1=\eta_3=0$}

The only nonvanishing Ricci rotation coefficients are $\tau_{1}, \tau_2, \tau_3$. 
We can suppose $\tau_1 \ne \tau_2$, since the $\tau_1=\tau_2$ case 
is obtained by the $\eta_1\to 0$ limit of (A-ii). 
In this case, $W_a^\pm =e_{2a} \pm \sqrt{(\tau_1+\tau_3)/(\tau_2-\tau_1)} e_{3a}$ 
is hypersurface-orthogonal, implying the existence of adapted coordinates. 
$W^\pm$ are real or complex conjugates. In either case, the derivation is similar to 
those described in section \ref{sec_[1,(11)]_case2_tau23>0} and \ref{sec_[1,(11)]_case2_tau23<0}.  
The final metric is written in a unified fashion as 
\begin{align}
\label{[1,11]metric:(A-i-1)}
e_1 =& \,-\left(\D t -t_0 \cos(\sqrt k r)\D \phi \right) \,, \qquad 
e_2=\cos (\sqrt\epsilon t/t_0) \D r-\frac{\sqrt\epsilon }{\sqrt k}\sin (\sqrt k r)\sin (\sqrt\epsilon t/t_0) \D \phi \,,\notag \\
e_3=& \,\sqrt{\epsilon \frac{\tau_1+\tau_3}{\tau_1-\tau_2}} 
\left(\frac{\sin (\sqrt\epsilon t/t_0)}{\sqrt\epsilon }\D r+\frac{1}{\sqrt k}\sin (\sqrt k r)\cos (\sqrt\epsilon t/t_0) \D \phi  \right)\,, 
\end{align}
where we have defined
\begin{align}
\label{}
k\equiv (\tau_1-\tau_2)(\tau_2-\tau_3) \,, \qquad t_0 \equiv -\frac{\epsilon }{\tau_1+\tau_3}\sqrt{\epsilon \frac{\tau_1+\tau_3}{\tau_1-\tau_2}} \,, \qquad 
\epsilon ={\rm sgn}\left(\frac{\tau_1+\tau_3}{\tau_1-\tau_2}\right)\,.  
\end{align}
The KVs are given by
\begin{align}
\label{}
K_1 =&\,
\sqrt k t_0 \cos(\sqrt \epsilon \phi) \,{\rm csc}(\sqrt k r)\partial_t +\sqrt \epsilon \sin(\sqrt \epsilon \phi) \partial_r +\sqrt k \cos(\sqrt \epsilon \phi) \cot (\sqrt k r)\partial_\phi \,, \\ 
K_2=&\, -\frac{\sqrt k}{\sqrt \epsilon} t_0 \sin(\sqrt \epsilon \phi)\, {\rm csc}(\sqrt k r)\partial_t + \cos(\sqrt \epsilon \phi) \partial_r -\sqrt k \epsilon^{3/2} \sin(\sqrt \epsilon \phi)\cot (\sqrt k r)\partial_\phi \,, \\ 
K_3=&\, \partial_\phi \,, 
\end{align}
The commutation relations are 
\begin{align}
\label{}
[K_1, K_2]= k K_3 \,, \qquad [K_2, K_3]=  K_1\,, \qquad 
[K_3, K_1]=\epsilon K_2 \,. 
\end{align}

\subsection{case (A-ii): $\kappa_2=\eta_3=\tau_2-\tau_1=0$ with $\eta_1\ne 0$}

The nonvanishing Ricci rotation coefficients are 
$\tau_1=\tau_2, \tau_3, \eta_1, \eta_2$, which are constrained to be 
\begin{align}
\label{[1,11]metric:(A-ii)constraints}
\tau_1=\frac{3\lambda_3}{\lambda_1-\lambda_2}\tau_3 \,, \qquad 
\eta_2=-\frac{\lambda_3-\lambda_1}{\lambda_2-\lambda_3}\eta_1 \,, \qquad 
3\lambda_3 \left(\frac{4\tau_3^2}{(\lambda_1-\lambda_2)^2}- \frac{\eta_1^2}{(\lambda_2-\lambda_3)^2 }\right)=1\,.
\end{align}
Since $[e_1,e_2]^a=0$, we can set $e_1{}^a=(\partial_t)^a$ and $e_2{}^a =(\partial_x)^a$. 
Equation (\ref{eq:app_comm_ortho}) can be integrated to give 
$e_3=[\eta_1t+(\tau_1-\tau_3)x+\omega_1(y)]\partial_t+[-(\tau_1+\tau_3)t-\eta_2 x +\omega_2(y)]\partial_x+\omega_3(y)\partial_y$, where 
$\omega_i$ are arbitrary functions of $y$. By $\int \D y /\omega_3 \to y$, one can set $\omega_3=1$, whereas  
$\omega_1$ and $\omega_2$  can be made to vanish by coordinate transformations $t\to t+g_1(y)$ and $x\to x+g_2(y)$ where 
$\eta_1 g_1+(\tau_1-\tau_3)g_2+\omega_1-g_1'=0$ and $(\tau_1+\tau_3)g_1+\eta_2 g_2 -\omega_2 +g_2'=0$. 
We thus arrive at 
\begin{align}
\label{[1,11]metric:(A-ii)}
e_1=-[\D t+(-\eta_1 t+(\tau_3-\tau_1)x)\D y]\,, \qquad 
e_2=\D x+[\eta_2 x+(\tau_1+\tau_3)t] \D y \,, \qquad 
e_3=\D y \,. 
\end{align}
The KVs are given by 
\begin{subequations}
\begin{align}
\label{}
K_1=&\, \partial_y \,, \\ 
K_2=& e^{\beta  y}\left[(\tau_1-\tau_3)  \cosh (\sqrt k y) \partial_t +\left(
-\frac 12 (\eta_1+\eta_2)\cosh(\sqrt k y)+\sqrt k \sinh(\sqrt ky)
\right)\partial_x \right] \,, \\
K_3=& e^{\beta  y}\left[\frac{(\tau_1-\tau_3)}{\sqrt k}\sinh(\sqrt k y)\partial_t 
+\left(\cosh(\sqrt ky)-\frac{\eta_1+\eta_2}{2\sqrt k}\sinh(\sqrt ky)\right)\partial_x \right]
\,,
\end{align}
\end{subequations}
where 
\begin{align}
\label{}
\beta  \equiv\frac{\eta_1-\eta_2}{2} \,, \qquad 
k\equiv\frac 14 (\eta_1+\eta_2)^2-(\tau_1^2-\tau_3^2) \,. 
\end{align}
The nonvanishing commutation relations are given by 
\begin{align}
\label{}
[K_1, K_2]=\beta   K_2+k K_3 \,, \qquad 
[K_3, K_1]=-K_2-\beta  K_3 \,. 
\end{align}

\subsection{case (B-i-1): $\eta_3=\tau_3-\tau_2=\eta_1=0$ with $\kappa_2 \ne 0$}

The nonvanishing Ricci rotation coefficients are 
$\kappa_2, \kappa_3 , \tau_1, \tau_2=\tau_3$. The constraints to be imposed on these 
parameters are 
\begin{align}
\label{}
\kappa_3=\frac{\lambda_1-\lambda_2}{\lambda_3-\lambda_1}\kappa_2 \,, \qquad 
\tau_2=-\frac{3\lambda_1 \tau_1}{\lambda_2-\lambda_3} \,, \qquad 
3 \lambda_1 \left(\frac{\kappa_2^2}{(\lambda_1-\lambda_3)^2}+\frac{4\tau_1^2}{(\lambda_2-\lambda_3)^2}\right)=1 \,. 
\end{align}
Now $[e_2 ,e_3]^a=0$, implying $e_2=\partial_x$ and $e_3=\partial_y$. 
$e_1$ can be inferred from (\ref{eq:app_comm_ortho}) as 
\begin{align}
\label{}
e_1=\omega_1(t) \partial_t +[-\kappa_2x+(\tau_1+\tau_2)y+\omega_2(t)] \partial_x +[(\tau_2-\tau_1)x
-\kappa_3 y+\omega_3(t) ] \partial_y \,. 
\end{align}
Without loss of generality, one can set $\omega_1 =1$ and $\omega_2=\omega_3=0$, yielding 
\begin{align}
\label{[1,11]metric:(B-i-1)}
e_1=-\D t\,, \qquad 
e_2 =\D x +[\kappa_2 x-(\tau_1+\tau_2)y ]\D t \,, \qquad 
e_3= \D y+[\kappa_3 y+(\tau_1-\tau_2)x]\D t \,. 
\end{align}
The KVs are given by 
\begin{subequations}
\begin{align}
\label{}
K_1=&\, \partial_t \,, \\
K_2 =& e^{\beta  t }\left[-\left(\frac 12(\kappa_3-\kappa_2)\cosh(\sqrt k t)
+\sqrt k \sinh(\sqrt k t)\right)\partial_x+(\tau_1-\tau_2)\cosh (\sqrt k t )\partial_y\right] \,,\\ 
K_3= &e^{\beta   t}\left[-\left(\cosh(\sqrt k t)+\frac {\kappa_3-\kappa_2}{2\sqrt k} \sinh(\sqrt k t)\right)\partial_x
+\frac{\tau_1-\tau_2}{\sqrt k}\sinh(\sqrt k t)\partial_y \right]  \,, 
\end{align}
\end{subequations}
satisfying 
\begin{align}
\label{}
[K_1, K_2]=\beta  K_2 +k K_3 \,, \qquad 
[K_3, K_1]= -K_2-\beta  K_3 \,, \qquad 
[K_2,K_3]=0 \,, 
\end{align}
where 
\begin{align}
\label{}
\beta  \equiv-\frac{\kappa_2+\kappa_3 }2\,, \qquad 
k\equiv\frac 14(\kappa_2-\kappa_3)^2-(\tau_1^2-\tau_2^2)
 \,.  
\end{align}

\section{Homogeneous metrics: Segre [(21)]} 
\label{sec_Segre [(21)]}

This section discusses the Segre type [(21)], in which the trace-free part of the Ricci tensor takes the form
\begin{align}
\label{Sab_[(21)]}
S_{ab}= S_{vv}u_a u_b \,, \qquad S_{vv}\ne 0 \,, 
\end{align}
where we have worked in a null frame (\ref{nullframe}). 
The Bianchi identity implies 
\begin{align}
\label{[(21)]:Bianchi}
\eta_u=0 \,, \qquad  \ma L_u S_{vv}+(2\kappa_u -\kappa_e)S_{vv}=0 \,. 
\end{align}
The following analysis is classified according to $\sigma=0$ or not, where
\begin{align}
\label{eq:sigma}
\sigma \equiv \ma L_e \varphi+\tau_v+\tau_e \,. 
\end{align}
Here and in what follows, we set $S_{vv}=e^{2\varphi} $. 
The above form (\ref{Sab_[(21)]}) of $S_{ab}$ is preserved under (\ref{null_localSO21_I}) and (\ref{null_localSO21_II}), which will 
be used in the following analysis. 

\subsection{$\sigma=0$, $\kappa_e\ne 0$}

The second obstruction matrix reads (4.36c) in \cite{Nozawa:2019dwu}, 
which vanishes iff
\begin{align}
\label{}
\{ \tau_v, \Sigma, \kappa_e e^\varphi , \eta_e e^{-\varphi} , \eta_v e^{-2\varphi} \}={\rm constants}  \,, 
\end{align}
where 
\begin{align}
\label{}
\Sigma \equiv \ma L_v (e^{-\varphi})+\kappa_v e^{-\varphi} \,. 
\end{align}
By virtue of $\kappa_e\ne 0$, 
the component $R_{ab}e^a u^b=0$ gives rise to $\tau_u=\tau_v$. 
From $\sigma=0$, $\Sigma={\rm constant}$ and Bianchi identity, 
one can derive the first-order equations for 
$\varphi$
\begin{align}
\label{[(21)]_varphieq}
\ma L_u \varphi=\frac 12 \kappa_e-\kappa_u  \,, \qquad 
\ma L_v\varphi =\kappa_v -\Sigma e^{\varphi} \,, \qquad 
\ma L_e \varphi=-(\tau_e+\tau_v) \,. 
\end{align}
From the Ricci identity  $R_{ab}u^a u^b=0$ we have $\ma L_u \kappa_e =\kappa_e(\kappa_e+\kappa_u)$. 
Inserting this into $\ma L_u (\kappa_ee^{\varphi})=0$ and using (\ref{[(21)]_varphieq}), we obtain $\kappa_e=0$, 
leading to the contradiction to the assumption $\kappa_e\ne 0$. 
This means that there exist no solutions in this class, which admit precisely 
3 KVs. 

Let us emphasize that this does not mean that this class fails to admit KVs more than 3. 
As illustrated in our previous paper \cite{Nozawa:2019dwu}, 
the class 3 Segre [(21)] allows solutions with 4 KVs.

\subsection{$\sigma\ne 0$}
\label{sec_Segre [(21)]_case2}

The second obstruction matrix is given by (4.37c) in \cite{Nozawa:2019dwu}, 
which vanishes if 
\begin{align}
\label{obstruction_Segre[(21)]_sigmanonzero}
 \ma L_i \left(\sigma-\frac 52 \tau_v\right)=0 \,, \qquad   \ma L_i \tau_v -\sigma^{-1} e^{\varphi}\kappa_e \ma L_i \Sigma  =0 \,, \qquad 
 \ma L_i (\kappa_e e^{\varphi})=0 \,, \qquad 
\Phi_i^{[(21)]}=\Theta_i ^{[(21)]}=0 \,, 
\end{align}
where 
\begin{align}
	%---
	\Phi_i ^{[(21)]}~\equiv~&
	\ma L_i \eta_v
	- 2 \eta_v \ma L_i \varphi
	+\frac{e^\varphi \ma L_i \ma L_v \Sigma}{\sigma}-\frac{e^\varphi (\ma L_i \varphi) (\ma L_v \Sigma)}{\sigma} \notag \\
	&
	+\Bigl(
	e^\varphi \ma L_e \Sigma
	-\ma L_v \sigma 
	+(\ma L_v \varphi -\kappa_v-\eta_e) \sigma
	\Bigr)
	\frac{e^\varphi \ma L_i \Sigma}{\sigma^2}\:, \\
	%---
	\Theta_i^{[(21)]} ~\equiv~& \ma L_i \eta_e
	-\eta_e \ma L_i \varphi
	-\frac{e^\varphi \ma L_i \ma L_e \Sigma}{\sigma}
	+
	\Bigl(
	e^{\varphi} \ma L_u \Sigma
	+ \ma L_e \sigma 
	+(\tau_u + \tau_v - \sigma)\sigma
	\Bigr) \frac{ e^\varphi \ma L_i \Sigma}{\sigma^2}\:.
\end{align}

From the Ricci identity $R_{ab}u^au^b=0$ and $  \ma L_u (\kappa_e e^\varphi)= 0 $, 
one must have $\kappa_e=0$. Since $\sigma$ and $\tau_v$ are now constant due to the first two conditions of (\ref{obstruction_Segre[(21)]_sigmanonzero}), the 
Ricci identity $S_{ab}u^av^b=S_{ab}e^a e^b=0$ gives rise to $R=-6 \tau_v^2$. 
$\Phi_e^{[(21)]}=0$ is now reduced to  
\begin{align}
\label{obst_[(21)]_phieq}
e^{2\varphi} =\frac{2(\sigma-2\tau_v)}{\sigma} 
[\eta_v\sigma^2 +\kappa_v^2(-\sigma+\tau_v)+\sigma \ma L_v \kappa_v +\kappa_v (\sigma-2\tau_v)\ma L_v \varphi
+\tau_v(\ma L_v \varphi)^2 -\sigma \ma L_v\ma L_v \varphi ]\,. 
\end{align}
Since $\tau_v=\frac 12 \sigma$ leads to the contradiction to $S_{vv}=e^{2\varphi} \ne 0$, 
we assume $\sigma \ne 2\tau_v $ in the hereafter. Then $\Phi_{u}^{[(21)]}=\Phi_v^{[(21)]} =0$ and 
$\Theta_i^{[(21)]}=0$ are automatically fulfilled. 
We have now three first-order system for $\ti \Sigma \equiv e^\varphi \Sigma$ as 
\begin{align}
\label{[(21)]_Sigmatildeeq}
\ma L_u \ti \Sigma=- \kappa_u \ti \Sigma - (\tau_u-\tau_v)\sigma \,, \qquad 
\ma L_v \ti \Sigma=\kappa_v \ti \Sigma-\eta_v\sigma  +\frac{\sigma}{2(\sigma-2\tau_v)} e^{2\varphi}-\frac{\tau_v}{\sigma}\ti \Sigma^2\,, \qquad 
\ma L_e \ti \Sigma=\eta_e \sigma-(\tau_e-\tau_v) \ti \Sigma \,. 
\end{align}
Since the compatibility of these equations are automatically satisfied, equation (\ref{obst_[(21)]_phieq}) is the 
unique constraint to be imposed. 

Let us note that all the conditions obtained above are invariant under (\ref{null_localSO21_I}). 
Making use of this freedom, one can choose $\tau_e+\tau_u=0$, 
permitting us to choose the triad (\ref{nullframe_uecommute}). 
The constancy of $\tau_v$ requires 
\begin{align}
\label{[(21)]_V3y}
\partial_y V_3~=~-2 \tau_v+\frac{\partial_z V_1+V_3 \partial_y V_1}{V_1}\,.
\end{align}
Inserting this into $S_{ab}e^b=0$, one finds a function $k_1(x)$ satisfying 
\begin{align}
\label{[(21)]_V2y}
\partial_y V_2~=~-k_1'(x) V_1+V_3 \left(-2\tau_v+\frac{\partial_z V_1}{V_1}\right)
+\frac{2V_2 \partial_y V_1}{V_1}+\partial_x V_1 \,.
\end{align}
From (\ref{[(21)]:Bianchi}) and (\ref{eq:sigma}), 
$\varphi $ is found to be 
\begin{align}
\label{[(21)]_varphi}
\varphi=(\sigma-2\tau_v)z+\log (k_2(x)V_1) \,,
\end{align}
where $k_2=k_2(x)$ is an arbitrary function of $x$. 
Comparison of  $S_{vv}=e^{2\varphi}$ in a coordinate basis with 
the one in (\ref{[(21)]_varphi}), we find a function 
$k_3=k_3(x,y)$ such that 
\begin{align}
\label{}
k_3(x,y)~=~& -\frac{\partial_x V_3}{V_1}+\frac{-2V_2+V_3^2}{V_1^3}\partial_z V_1 -\frac{V_3}{V_1^2}\partial_z V_3 
+\frac{V_3}{V_1^2}\partial_x V_1+\frac{\partial_z V_2}{V_1^2}\notag \\
& +\frac{V_3}{V_1^2}(-k_1'(x) V_1-\tau_v V_3)+\frac{2\tau_v V_2}{V_1^2}-\frac{e^{2(\sigma-2\tau_v)}}{2(\sigma-2\tau_v)}k_2^2 \,, 
\end{align}
which can be rewritten into 
\begin{align}
\label{}
\partial_x \left[\frac{V_3 e^{2\tau_v z+k_1}}{V_1}\right]  =\partial_z \left[
e^{2\tau_v z+k_1}\left(\frac{2V_2-V_3^2}{2V_1^2}-\frac{k_3}{2\tau_v}-
\frac{e^{2(\sigma-2\tau_v) z}}{4(\sigma-2\tau_v)(\sigma-\tau_v)}k_2^2 \right)
\right]\,.
\end{align}
This implies the existence of a function $F=F(x,y,z) $ such that the terms in the square bracket on the left-hand side is $\partial_z F$
and the terms in the square bracket on the left-hand side is $\partial_x F$, which amounts to 
\begin{subequations}
\label{[(21)]_V23F}
\begin{align}
V_2 =& \frac 14 V_1^2 \left[
\frac{e^{2(\sigma-2\tau_v)z}k_2^2}{(\sigma-2\tau_v)(\sigma-\tau_v)}
+\frac{2k_3}{\tau_v} +2 e^{-2(2\tau_v z+k_1)}[(\partial_zF)^2+2 e^{2\tau_v z+k_1} \partial_x F ] 
\right]  \,, \\
V_3=& e^{-2\tau_v z-k_1 } V_1 \partial_z F \,. 
\end{align}
\end{subequations}
From the compatibility of (\ref{[(21)]_V3y}), (\ref{[(21)]_V2y}) and (\ref{[(21)]_V23F}), 
we have 
\begin{align}
\label{}
V_1 =\frac{e^{2\tau_v z+k_1(x)} }{k_4(y)-\partial_y F }\,, \qquad 
k_3=k_3 (x) \,. 
\end{align}
where $k_4$ is a function of $y$. 
$\ti \Sigma$ is now computed to yield
\begin{align}
\label{}
\ti \Sigma =-\frac{e^{2z\tau_v+k_1(x)}(k_2(x)k_1'(x)+k_2'(x))+\sigma k_2(x)\partial_z F}{k_2(x)(k_4(y)-\partial_y F)} \,. 
\end{align}
The obstruction (\ref{[(21)]_Sigmatildeeq}) reduces to 
\begin{align}
\label{[(21)]_Sigmatildeeq2}
(\sigma+2\tau_v)k_2'(x)^2+k_2(x)^2 [\sigma^2k_3(x)+(-\sigma+\tau_v)k_1'(x)^2-\sigma k_1''(x)]
-k_2(x)[(\sigma-2\tau_v) k_1'(x)k_2'(x)+\sigma k_2''(x)]=0\,. 
\end{align}

Defining $\hat y=\int k_4(y) \D y -F$, $\hat x= \int e^{-k_1(x)} \D x$,
the metric now reads
\begin{align}
\label{}
\D s^2 =2 e^{-2\tau_v z}\D \hat x \D \hat  y+ \D z^2 - \D \hat x^2 \left[
\frac{e^{2(\sigma-2\tau_v)z}}{2(\sigma-2\tau_v)(\sigma-\tau_v)}\hat k_2(\hat x)^2+\frac{\hat k_3(\hat x)}{\tau_v}
\right] \,, 
\end{align}
where 
$\hat k_{2}(\hat x)=k_{2}(x)e^{k_1(x)}$ and 
$\hat k_{3}(\hat x)=k_{3}(x)e^{2k_1(x)}$. 
A further coordinate change 
$\ti y= \hat y-\frac{e^{2\tau_v z}}{2\tau_v} h_1'(\hat x)$, 
$\ti z=z+h_1(\hat x)$, 
$\ti x =\int e^{2\tau_v  h_1(\hat x)} \D \hat x$
is used to simplify the metric as 
\begin{align}
\label{}
\D s^2=2 e^{-2\tau_v \ti z}\D \ti x \D \ti  y+ \D \ti z^2-\frac{\ti k_2 (\ti x)^2 }{2(\sigma-2\tau_v)(\sigma-\tau_v)}
e^{2(\sigma-2\tau_v)\ti z}\D \ti x^2 \,, 
\end{align}
where 
$\ti k_2 (\ti x)= e^{-2\tau_v h_1(\hat x)}\hat k_2(\hat x) $ and 
$h_1$ has been chosen to satisfy $h_1''(\hat x)=\hat k_3 (\hat x)+\tau_v h_1'(\hat x)^2$. 
In this coordinate system, 
the last obstruction condition (\ref{[(21)]_Sigmatildeeq2}) becomes 
\begin{align}
\label{[(21)]_Phie}
( \sigma+\tau_v )\ti k_2'(\ti x)^2 -\sigma \ti k_2(\ti x)\ti k_2''(\ti x)=0 \,. 
\end{align}
The solution to this differential equation branches into 
two cases, depending on $\tau_v =0$ or not. 

When $\tau_v \ne 0$, the solution to (\ref{[(21)]_Phie}) is 
$\ti k_2(\ti x)=c_0 (-\tau_v \ti x)^{-\sigma/\tau_v}$. 
One can perform the transformation 
$\ti x=-1/(\sigma^2 x)$, $\ti y=y+\frac{e^{2\tau_v z}}{2\tau_v^2 x}$ and 
$\ti z=z-\frac 1{\tau_v}\log (\sigma x)$ to bring the metric into  the homogeneous plane wave
\begin{align}
\label{[(21)]metric}
\D s^2 =2 e^{-2\tau_v z}\D x \D y+ \D  z^2 +C_0 e^{2(\sigma-2\tau_v)z}\D x^2 \,,
\end{align}
where $C_0$ is a constant. On the other hand, 
the solution to (\ref{[(21)]_Phie}) is 
$\ti k_2(\ti x)=c_1 e^{c_2 \ti x}$ for $\tau_v=0$. By 
$\ti x\to x$, $\ti y=y-\frac{c_2^2}{2\sigma^2} x+\frac{c_2}{\sigma}z$, 
$\ti z=z-\frac{c_2}{\sigma}z$, the metric can be written again into the form (\ref{[(21)]metric}) 
with $\tau_v=0$. 
In either case, three KVs are 
\begin{align}
\label{}
K_1=\partial_x \,, \qquad K_2= \partial_y \,, \qquad K_3=-\frac{\sigma-2\tau_v}{\sigma}
x\partial_x +y \partial_y +\sigma^{-1} \partial_z
\end{align}
with 
\begin{align}
\label{}
[K_1, K_3]=-\frac{\sigma-2\tau_v}{\sigma } K_1 \,, \qquad [K_2, K_3]=K_2\,. 
\end{align}
For $\sigma=\frac 32\tau_v$, the above [(21)] metric (\ref{[(21)]metric}) is conformally flat.

\section{Homogeneous metrics: Segre [21]} 
\label{sec_Segre[21]}

For the Segre [21] type, the trace-free Ricci tensor reads 
\begin{align}
\label{Sab_[21]}
S_{ab}= 2\lambda u_{(a}v_{b)} +S_{vv}u_a u_b -2\lambda e_a e_b \,, 
\end{align}
where the class 3 condition amounts to $\lambda={\rm const}(\ne 0)$. 
From the Bianchi identity, we have 
\begin{align}
\label{Bianchi_[21]}
\kappa_e =0 \,, \qquad \eta_u =-3\lambda e^{-2\varphi}(\tau_u+\tau_v) \,, \qquad 
\ma L_u \varphi=-\kappa_u +\frac 32 e^{-2\varphi} \eta_e \lambda \,, 
\end{align}
where $S_{vv}=e^{2\varphi}$. 
The second obstruction matrix is given by (4.41a) in \cite{Nozawa:2019dwu}, 
which vanishes iff 
\begin{align}
\label{}
\{\tau_u, \tau_v, \sigma , \Sigma, \eta_v e^{-2\varphi} , \eta_e e^{-\varphi} \}={\rm constants} \,.
\end{align}

In the Jordan basis (\ref{Sab_[21]}), 
the Ricci identity (\ref{eq:app_ricci_null}) requires
\begin{align}
\label{[21]_Ricciidcond}
( 2\sigma+\tau_u-\tau_v)(\tau_u+\tau_v)=0 \,, \qquad \Sigma (\tau_u+\tau_v)=0 \,. 
\end{align}
From the Bianchi identity, definition of $\Sigma$ and $\sigma$ give rise to the 
1st-order system 
\begin{align}
\label{}
\ma L_u \varphi=-\kappa_u +\frac 32 e^{-2\varphi}\eta_e \lambda \,, \qquad 
\ma L_v \varphi=\kappa_v -\Sigma e^\varphi \,, \qquad 
\ma L_e \varphi= \sigma-\tau_e-\tau_v \,. 
\end{align}
The compatibility conditions (\ref{eq:app_comm_null}) for these equations give 
\begin{subequations}
\label{Segre[21]_phicompeq}
\begin{align}
0=&\, (\tau_v-\tau_u )\sigma -3  \lambda [1+e^{-\varphi} \eta_e \Sigma +2 \eta_v(\tau_u+\tau_v)e^{-2\varphi} ]\,, \\
0=&\, 2 (\tau_u+\tau_v)e^{\varphi} \Sigma +\eta_e (\sigma+3\tau_u+\tau_v) \,, \\ 
0=&\, -\frac 32 e^{-2\varphi} \eta_e \eta_v \lambda +\eta_e \sigma -(\sigma-2\tau_v) e^\varphi \Sigma \,. 
\end{align}
\end{subequations}
Inspecting (\ref{[21]_Ricciidcond}), the following analysis falls into two cases, according to the vanishing of $\tau_u+\tau_v$.

\subsection{$\tau_u+\tau_v\ne 0$}

From (\ref{[21]_Ricciidcond}), 
we obtain
\begin{align}
\label{}
\Sigma =0 \,, \qquad \sigma =\frac 12 (\tau_v -\tau_u) \,. 
\end{align}
From (\ref{Segre[21]_phicompeq}), we must have $\eta_e=0$. 
Together with $\ma L_i (\eta_ve^{-2\varphi})=0$, the rest of the Ricci identity 
demands 
\begin{align}
\label{}
\tau_u =\tau_v \,, \qquad e^{2\varphi}=-4\eta_v \tau_v \,,
\end{align}
leading to $\sigma=0$. 

Since we have extracted all the conditions to be imposed, we move on to introduce local coordinates. 
To this aim, 
let us consider the rescaled frame
\begin{align}
\label{[21]_rescale}
\hat u^a =e^{\varphi} u^a \,, \qquad \hat v^a =e^{-\varphi} v^a \,, 
\end{align}
Substituting this into (\ref{eq:app_comm_null}), one finds 
\begin{align}
\label{[21]_hatcom}
[ \hat u, \hat v]^a =0 \,, \qquad 
[e, \hat u]^a=-6\lambda \tau_v \hat v^a \,, \qquad 
[e, \hat v]^a =-\frac 1{4\tau_v}\hat u^a +2\tau_v \hat v^a \,. 
\end{align}
The first equation implies that we can employ the coordinate system in such a way that 
$\hat u^a$ and $\hat v^a$ are coordinate vectors $\hat u^a= (\partial_y)^a$, 
$\hat v^a=(\partial_x )^a$. The second and third equations in (\ref{[21]_hatcom})
are integrated to yield 
\begin{align}
\label{}
e^a =\left[6\lambda \tau_v y-2\tau_v x+\omega_1(z) \right] (\partial_x)^a 
+\left(\frac{x}{4\tau_v}+\omega_2(z) \right)(\partial_y)^a +\omega_3(z) (\partial_z)^a \,, 
\end{align}
where $\omega_{1,2,3}(z)$ are arbitrary functions of $z$. 
Since (\ref{[21]_rescale}) corresponds just to  the rescaling of the frame (\ref{null_localSO21_I}), 
one can set $\varphi=0$ without losing any generality.  
By $z\to  \int \omega_3^{-1}\D z$, we can achieve $\omega_3(z)=1$. 
A change of variables $x\to  x+g_1(z)$ and $y\to y+g_2(z)/(4\tau_v)$ allows one to 
set $\omega_{1,2}=0$ by choosing $g_{1,2}$ to satisfy 
$g_1+4\tau_v \omega_2-g_2'=0$ and 
$2\tau_v g_1 -\frac 32 \lambda g_2-\omega_1 +g_1'=0$. 
Then, the triad frame simplifies to 
\begin{align}
\label{}
u_a \D x^a =\D x +2 \tau_v (x-3\lambda y)\D z \,, \qquad 
v_a \D x^a=\D y-\frac{x}{4\tau_v} \D z \,, \qquad e_a \D x^a=\D z \,. 
\end{align}
We therefore arrive at 
\begin{align}
\label{}
\D s^2=2 [\D x+2(x-3\lambda y)\tau_v \D z ] \left(\D y-\frac{x}{4\tau_v} \D z\right)+\D z^2 \,. 
\end{align} 
The KVs are given by
\begin{subequations}
\label{[21]_KV_case1}
\begin{align}
K_1=&\, \partial_z \,, \\ 
K_2=&\,  e^{-\tau_v z}\left[-4 \tau_v \left(\tau_v \cosh (\sqrt k z)-\sqrt k \sinh(\sqrt kz)\right)\partial_x 
+ \cosh (\sqrt kz) \partial_y\right]  \,, \\ 
K_3= &\, e^{-\tau_v z}\left[4 \tau_v\left(\cosh (\sqrt kz)-\frac{\tau_v}{\sqrt k}\sinh(\sqrt k z)\right)\partial_x
+\frac{1}{\sqrt k}\sinh (\sqrt k z)\partial_y\right]\,, 
\end{align}
\end{subequations} 
with the commutators
\begin{align}
\label{[21]_KV_case1_algebra}
[K_1, K_2]= k K_3 -\tau_v K_2 \,, \qquad 
[K_1, K_3]=K_2-\tau_v K_3 \,. 
\end{align}
Here we have defined $k=\frac 32 \lambda +\tau_v^2$. 
The expressions in (\ref{[21]_KV_case1}) are valid in either sign of $k$.

\subsection{$\tau_u+\tau_v=0$}

In this case, we have $\eta_u=0$ from the Bianchi identity (\ref{Bianchi_[21]}). 
The Ricci identity in the Jordan basis (\ref{Sab_[21]}) demands
\begin{align}
\label{[21]_case1_Ricid}
\eta_e=0\,, \qquad e^{2\varphi}=2\eta_v (\sigma-\tau_v) \,. 
\end{align} 
The compatibility conditions (\ref{Segre[21]_phicompeq}) give 
\begin{align}
\label{Segre[21]_phicompeq2}
\lambda =\frac 23 \sigma \tau_v \,, \qquad 
(\sigma-2\tau_v ) \Sigma =0 \,. 
\end{align}

Let us now introduce the local coordinates. We have the gauge freedom  (\ref{null_localSO21_I}), 
which preserves all the above conditions on connections and curvatures. This allows us to work in a gauge 
$\tau_e+\tau_u=0$, whence we can  employ the triad (\ref{nullframe_uecommute}). 
Requiring $\eta_e=0$, we have $\partial_z (\log V_3)=\partial_z (\log V_1)$. 
$\tau_u+\tau_v=0$ is solved as $V_1=(\partial_y F(x,y))^{-1}$. Inserting into the definition of $\tau_v(={\rm const})$, 
we get $V_3 =(-2\tau_v F+\omega_{1}(x) )/\partial_y F$. From $S_{ab}v^a e^b=0$, 
one can derive $\partial_z V_2=f_1(x,z)/(\partial_y F)^2$. 
The constancy of $\sigma $ and the second equation of (\ref{[21]_case1_Ricid}) asks for 
$f_1(x,z)= e^{2(\sigma-2\tau_v)z} f_{11}(x)$. From $\Sigma={\rm const.}$ condition, 
we obtain 
\begin{align}
\label{}
V_2 =\frac{1}{(\partial_y F)^2}&\Biggl[ \frac 12 F\partial_x (\log f_{11})+
\tau_v(2\tau_v-\sigma)F^2 +V_{21}(x,z) -\partial_x F \notag \\
&
+F\left(\sqrt{2(\sigma-\tau_v)e^{2(\sigma-2\tau_v)z}f_{11} }\Sigma +(\sigma-2\tau_v)\omega_{1}\right)
\Biggr]\,, 
\end{align}
where $V_{21}(x,z) $ is a function independent of $y$. 
Comparing this expression with  $\partial_z V_2=f_1(x,z)/(\partial_y F)^2$, we get 
\begin{align}
\label{[21]_V2comp}
\partial_z V_{21}(x,z) =e^{2(\sigma-2\tau_v)z} f_{11}(x) \,. 
\end{align}

Redefining $F(x,y)\to y$, the resulting metric is cast into
\begin{align}
\label{}
\D s^2= &\, 2 \D x \D y + [\D z+(2\tau_v y -\omega_{1}(x) )\D x ]^2 
\notag \\ 
&  -2
\left[\tau_v(2\tau_v-\sigma)y^2 +V_{21}(x,z) 
+y \left(\sqrt{2(\sigma-\tau_v)e^{2(\sigma-2\tau_v)z}f_{11}(x) }\Sigma +\frac {f_{11}'(x)}{2f_{11}(x)}
+(\sigma-2\tau_v)\omega_{1}(x)\right) \right]\D x^2 
\,. \label{[21]_metric0}
\end{align}
Our remaining task is to impose the second condition in (\ref{Segre[21]_phicompeq2})
and (\ref{[21]_V2comp}). In view of (\ref{Segre[21]_phicompeq2}), this will be implemented separately depending on $\sigma=2\tau_v$ or not. 

\subsubsection{$\sigma=2 \tau_v$}

Solving (\ref{[21]_V2comp}), we have $V_{21}(x,z)=f_{11}(x)z+\omega_2(x)$. 
Let us consider the following coordinate transformations 
\begin{align}
\label{[21]_CT}
x\to h( x)\,, \qquad y\to \frac{y-g_1(x)}{h'(x)} \,, \qquad 
z \to z+g_2(x)\,, 
\end{align}
and choose functions $h$ and $g_{1,2}$ to satisfy 
\begin{align}
\label{}
&f_{11}(h(x)) h'(x)^2 =C_1\,, \qquad 
\omega _1(h(x)) h'(x)=g_2'(x) -2\tau_v g_1(x) \,, \notag \\ 
&C_1 \omega_2 (x)+f_{11}(x)(-\sqrt{2C_1\tau_v}\Sigma g_1(x)+C_1 g_2(x) +g_1'(x))=0 \,, 
\end{align}
where $C_1$ is a nonvanishing constant with ${\rm sgn}(C_1\tau_v) \ge 0$. This procedure amounts to set 
$f_{11}(x)=C_1$ and $\omega_1(x)=\omega_2(x)=0$. 
The metric (\ref{[21]_metric0}) is then brought into 
\begin{align}
\label{[21]_metric_case1}
\D s^2=2 \D x \left[\D y-\left(C _1z +\sqrt{2\tau_v C_1}\Sigma y \right)\D x\right]
+(\D z+2\tau_vy \D x)^2 \,. 
\end{align}
KVs are given by
\begin{subequations}
\begin{align}
\label{[21]_KV_case2_type1}
K_1=&\, \partial_x\,,\\ 
K_2=&\,  e^{\Sigma_* x} \left[ -\left(\Sigma_* \cosh (\sqrt k x)+\sqrt k \sinh (\sqrt k x)\right)\partial_y
+2  \tau_v \cosh (\sqrt k x) \partial_z\right] \,, \\
K_3=&\,  e^{\Sigma_* x} \left[-\left( \cosh(\sqrt k x)+\frac{\Sigma_*}{\sqrt k}\sinh(\sqrt k x)\right)\partial_y 
+\frac{2 \tau_v }{\sqrt k} \sinh (\sqrt k x) \partial_z\right] \,, 
\end{align}
\end{subequations}
where we have denoted 
\begin{align}
\label{}
\Sigma_*=\sqrt{\frac{C_1\tau_v}2}\Sigma \,, \qquad 
k=\frac 12 \tau_v C_1 (-4 +\Sigma^2) \,. 
\end{align}
The KVs in (\ref{[21]_KV_case2_type1}) are well-defined in either sign of $k$ and obey the 
following commutation relations 
\begin{align}
\label{}
[ K_1, K_2]=  k K_3+\Sigma_* K_2\,, \qquad 
[K_1, K_3]=\Sigma_* K_3 +K_2 \,. 
\end{align}
This algebra is isomorphic to (\ref{[21]_KV_case1_algebra}).

\subsubsection{$\sigma\ne 2\tau_v $}

Solving (\ref{[21]_V2comp}), we get 
$V_{21}(x,z)=\frac 12 \sigma_*^{-1} e^{2\sigma_*z}f_{11}(x)+\omega_3(x)$, 
where $\sigma_*\equiv \sigma-2\tau_v $ is a constant. Imposing $\Sigma=0$,  we perform the coordinate transformations (\ref{[21]_CT}) and 
choose functions $h$, $g_{1,2}$ to satisfy 
\begin{align}
\label{}
&  \omega_1(h(x))h'(x)=g_2'(x) -2\tau_v g_2(x)\,, \qquad 
2\sigma_* g_2(x)+\log \left(\frac{f_{11}(h(x))h'(x)^2}{2C_2 \sigma_*}\right)=0 \,, \notag \\
& \sigma_* \tau_v g_1(x)^2 +\frac{2C_2}{ f_{11}(h(x))} e^{-2\sigma_* g_2 (x) } \sigma_*\omega_3(x)+g_1'(x)=0 \,,
\end{align}
where $C_2$ is a nonvanishing constant. 
Then, the metric is transformed into 
\begin{align}
\label{metric[21]_caseBsigmanetauv}
\D s^2=2\D x \left[\D y +\left(-C_2 e^{2\sigma_* z} + \sigma_* \tau_v y^2 \right)\D x \right]+(\D z+2 \tau_v y \D x)^2 \,, 
\end{align}
The KVs are given by
\begin{align}
\label{}
K_1=& \partial_x \,, \qquad 
K_2= x\partial_x-y \partial_y -\sigma_*^{-1} \partial_z \,, \qquad 
K_3=  \tau_v x^2\partial_x +\left(-2\tau_vxy +\frac{1}{\sigma_* }\right)\partial_y -\frac{2\tau_vx}{\sigma_*} \partial_z \,, 
\end{align}
satisfying $\mathfrak{so}(1,2)$ algebra
\begin{align}
\label{}
[K_1, K_2]=K_1\,, \qquad 
[K_2, K_3]=K_3 \,, \qquad 
[K_3,K_1]=-2\tau_v K_2\,. 
\end{align}
When $\tau_v=0$, the results in section \ref{sec_Segre [(21)]_case2} is recovered.

\section{Homogeneous metrics: Segre [3]} 
\label{sec_Segre [3]}

In the Segre [3] case, the trace-free Ricci tensor can be cast into 
\begin{align}
\label{}
S_{ab}= S_{vv}u_a u_b +2  u_{(a}e_{b)}\,, 
\end{align}
where we have exploited the rescaling freedom (\ref{null_localSO21_I}) 
to fix $S_{ve}=1$. 
The Bianchi identity puts the following restrictions
\begin{align}
\label{}
\eta_u=0 \,, \qquad \kappa_u=2\kappa_e \,, \qquad 
\ma L_u S_{vv}+3\kappa_e S_{vv}=2\tau_u+\tau_v-\tau_e \,. 
\end{align}
The second obstruction matrix boils down to (4.50a) in \cite{Nozawa:2019dwu}, 
which vanishes iff 
\begin{align}
\label{[3]_cond}
\{\kappa_e, \tau_e-3\tau_v , \kappa_e S_{vv}+2\tau_v\}={\rm constants} \,, \qquad 
\Phi_i^{[3]}=\Xi_i^{[3]}=\Theta_i ^{[3]}=0\,. 
\end{align}
Here 
\begin{subequations}
\begin{align}
\label{}
\Phi_i^{[3]}  \equiv\, & \ma L_i \kappa_v+\frac{\tau_v+\tau_e}2 \ma L_i S_{vv}\,, \label{Phi_[3]} \\ 
   \Xi_i ^{[3]} \equiv\, &\ma L_i \eta_e +\frac 12 \ma L_i \ma L_e S_{vv}+\frac 14 (\tau_v+\tau_e-3\kappa_e S_{vv})\ma L_i S_{vv} \,,  \\
   \Theta_i^{[3]} \equiv\,  &\ma L_i \eta_v- \frac 12 \ma L_i \ma L_v S_{vv}+\frac 14 (\ma L_e S_{vv}+2\kappa_v+2\eta_e)\ma L_i S_{vv} \,.
\end{align}
\end{subequations}

From $S_{uu}=0$, we have $\kappa_e=0$, leading to the conclusion that 
$\tau_e$ and $\tau_v$ are constants as a result of (\ref{[3]_cond}). 
A simple computation  yields
\begin{align}
\label{}
\Phi_u^{[3]} =-\frac 12(\tau_e+\tau_v)(\tau_e-3\tau_v) \,, \qquad 
\Theta_u^{[3]}  =-1-\frac 14 (\tau_e-3\tau_v ) (2\eta_e +\ma L_e S_{vv}) \,. 
\end{align}
It follows that $\Phi_u^{[3]} =\Theta_u^{[3]} =0$ are simultaneously satisfied only for $\tau_e=-\tau_v$. 
$\Phi_e^{[3]}=0$ gives rise to $2\kappa_v \tau_v=1$. 
Other components of $\Xi_i^{[3]}=\Theta_i^{[3]}=0$ are reduced to 
\begin{align}
\label{[3]_Svveq}
\ma L_u S_{vv}=2(\tau_u+\tau_v) \,, \qquad 
\ma L_v S_{vv}=\frac {S_{vv}+2\kappa_v^2+2\eta_v\tau_v}{\tau_v} \,, \qquad 
\ma L_e S_{vv}=2(\kappa_v -\eta_e) \,, 
\end{align}
whose integrability is automatically assured. 

Using the type II gauge transformations (\ref{null_localSO21_II}), one can set $\tau_u+\tau_e=0$, 
while all the above conditions are kept invariant. 
In this gauge, one sees $[u, e]^a=0$, allowing us to employ the frame (\ref{nullframe_uecommute}). 
$\kappa_u=0$ implies $V_1=V_1(x,z)$, whereas $\tau_e+\tau_v=0$ is solved to give 
$V_3=\partial_z V_{31}(x,z)$. The constancy of $\tau_v$ and $\kappa_v=(2\tau_v)^{-1}$ gives rise to $V_{1}=e^{2\tau_v z} H(x)$
and $V_2 =\frac{y}{2\tau_v}+V_{21}(x,z)$. 
Inserting these expressions into 
the first and the third equations in (\ref{[3]_Svveq}), one can conclude 
$S_{vv}=4\tau_v y+\frac{z}{\tau_v}-4\tau_v V_{31}(x,z)+S_0(x)+2 \partial_z V_{31}(x,z)$. 
Substituting next into the second equation in (\ref{[3]_Svveq}), one can integrate in $z$ to derive
\begin{align}
\label{}
V_{21}(x,z) = \frac{1}{32\tau_v^4}&\Bigl[
3+4\tau_vz+4\tau_v^2 \Bigl\{S_0(x) +2\tau_v\Bigl(-2 e^{4\tau_v z} \tau_v k_1(x) -2 V_{31}(x,z) +2 \tau_v(\partial_zV_{31}(x,z))^2 
\notag \\
& - e^{2\tau_vz}
H(x) (S_0'(x) -4\tau_v \partial_x V_{31}(x,z) )\Bigr)\Bigr\}
\Bigr] \,. 
\end{align}
Replacing $x\to \int \frac{\D x}{H(x)}$, 
$y\to y-V_{31}(x,z) +\frac{S_0(x)}{4\tau_v}+\frac{3}{16 \tau_v^3}$, one can elicit 
\begin{align}
\D s^2= 2 \D x \left[e^{-2\tau_v z} \D y +\left\{- \left(\frac{y}{2\tau_v}+\frac{z}{8\tau_v^3} \right)e^{-4\tau_v z} +\frac 12 k_1(x) \right\}\D x\right]+\D z^2 \,. 
\end{align}
By the extra coordinate transformations 
\begin{align}
\label{}
x=g_1(\hat x)\,, \qquad y=\hat y-\frac{\log (g_1'(\hat x))}{8\tau_v^2}-\frac{e^{2\tau_v \hat z}g_1''(\hat x)}{4\tau_v^2g_1'(\hat x)} \,, \qquad z=\hat z+\frac{1}{2\tau_v}\log (g_1'(\hat x)) \,,  
\end{align}
the metric can be brought into 
\begin{align}
\label{}
\D s^2= 2 \D \hat x \left[e^{-2\tau_v \hat z} \D \hat y +\left\{- \left(\frac{\hat y}{2\tau_v}+\frac{\hat z}{8\tau_v^3}  \right)e^{-4\tau_v \hat z}+\frac 12 k_1(g_1(\hat x)) g_1'(\hat x)^2-\frac 1{2\tau_v^2}(Sg_1)(\hat x) \right\}\D \hat x\right]+\D \hat z^2 \,,  
\end{align}
where $(Sg_1)$ denotes the Schwarzian derivative of $g_1$, 
\begin{align}
\label{}
(Sg_1)(\hat x)=\frac{g_1'''(\hat x)}{g_1'(\hat x)}-\frac 32 \left(\frac{g_1''(\hat x)}{g_1'(\hat x)}\right)^2 \,. 
\end{align}
This allows us to set $k_1(x)=0$, i.e., 
\begin{align}
\label{metric_[3]}
\D s^2= 2 \D x \left[e^{-2\tau_v z} \D y - \left(\frac{y}{2\tau_v}+\frac{z}{8\tau_v^3} \right)e^{-4\tau_v z} \D x\right]+\D z^2 \,. 
\end{align}
The KVs are therefore given by
\begin{align}
\label{}
K_1=\partial_x \,, \qquad 
K_2= x \partial_x -\frac{1}{8\tau_v^3} \partial_y +\frac{1}{2\tau_v}\partial_z \,, \qquad 
K_3= x^2 \partial_x -\frac{x+2 \tau_v e^{2\tau_vz}}{4 \tau_v^3}\partial_y +\frac{x}{\tau_v}\partial_z \,, 
\end{align}
with $\mathfrak{so}(1,2)$ algebra
\begin{align}
\label{}
[K_1,K_2]=K_1 \,, \qquad 
[K_1, K_3]= 2 K_2 \,, \qquad 
[K_2, K_3]=K_3 \,. 
\end{align}

Since none of the known stress-energy tensors is of the Segre [3] form, 
no particular attention has been focused on this class of metrics. 
Recently, the authors in \cite{Martin-Moruno:2018coa} discussed the construction for the 
metrics in Segre [3]. It is however suspicious whether the metric 
(\ref{metric_[3]}) is sourced by physical stress-energy tensor.

\section{Homogeneous metrics: Segre [$z\bar z1$]} 
\label{sec_Segrezz*1}

Lastly, we shall investigate the Segre [$z\bar z1$] class, for which the trace-free Ricci tensor 
$S^a{}_b$ obeys the following eigensystem 
\begin{align}
\label{}
S^a{}_b j^b_\pm  =\lambda _\pm j_\pm^a \,, \qquad S^a{}_b j^b = \lambda j^a \,, 
\end{align}
where $\lambda _\pm $ are complex eigenvalues corresponding to 
the complex eigenvectors $j_+^a =(j_-^a)^*$, $j^a$ is spacelike and $\lambda_++\lambda_- +\lambda=0$. 
Denoting $j_\pm ^a = e_1{}^a \pm i e_2{}^a $ and $\lambda_\pm =\alpha \pm i \beta $ ($\beta \ne 0$), 
$S_{ab}$ then takes the form 
\begin{align}
\label{zzb1_Segre}
S_{ab}=\alpha (-e_{1a}e_{1b}+e_{2a}e_{2b})+2\beta  e_{1(a}e_{2b)} -2\alpha e_{3a}  e_{3b} \,. 
\end{align}
Here 
$\{e_i\}$ denotes the orthonormal frame with $\varepsilon=-1$. 
From the class 3 condition, 
the eigenvalues ($\alpha, \beta$) are constants. 

The second obstruction matrix reads (4.59) in \cite{Nozawa:2019dwu}, which vanishes 
provided 
\begin{align}
\label{}
\{\kappa_3, \eta_1, \eta_2, \eta_3, \tau_2 , \tau_3 \} ={\rm constants} \,.  
\end{align}
Through the Bianchi identity, the Ricci rotation coefficients are related by 
\begin{align}
\label{zzb1_Bia}
\eta_3-2\kappa_1=\frac{3\alpha}{\beta} \kappa_3 \,, \qquad 
\kappa_3+2\kappa_2+\frac{3\alpha}{\beta}\eta_3=0\,, \qquad 
\tau_1+\tau_2=\frac{3\alpha}{\beta}(\eta_1-\eta_2) \,. 
\end{align}
Since  $\{\kappa_1, \kappa_2 , \tau_1\}$ are also constants, 
we deduce the homogeneity of the solution. 

Let us first suppose $\alpha=0$.  
Solving (\ref{zzb1_Bia}) with respect to ($\kappa_3, \eta_3, \tau_2$) and 
 inserting into $2S_{11}=S_{33}$, we get 
 $\eta_1^2-\eta_2^2=6 (\kappa_1^2+\kappa_2^2)$, allowing us to parameterize
 \begin{align}
\label{}
 \eta_1 =\sqrt 6 \kappa \cosh \phi_1 \,, \qquad \eta_2 = \sqrt 6 \kappa \sinh \phi_1\,, \qquad 
 \kappa_1 =\kappa \cos \phi_2 \,, \qquad \kappa_2=\kappa \sin\phi_2 \,, 
\end{align}
where $\kappa=\sqrt{\kappa_1^2+\kappa_2^2}$. 
With this parameterization, 
one can derive  $\kappa=0$ from $S_{31}=S_{32}=0 $, which gives rise to 
$S_{12}=0$. This is a contradiction to the Segre [$z\bar z 1$] condition ($\beta\ne 0$). 
In the hereafter, we therefore assume $\alpha \ne 0$. 

Solving (\ref{zzb1_Bia}) with respect to ($\kappa_3, \eta_2, \eta_3$) 
and plugging into the Ricci identity $S_{12}=S_{21}$, one ends up with 
$(9\alpha^2+\beta^2)(\tau_2^2-\tau_1^2)=18\alpha^2(\kappa_1^2+\kappa_2^2)$. 
We can thus parametrize 
\begin{align}
\label{}
\tau_2 =\tau\cosh \phi_1\,, \qquad 
\tau_1 =\tau \sinh \phi_1 \,, \qquad 
\kappa_1 =\kappa \cos \phi_2 \,, \qquad 
\kappa_2 =\kappa \sin \phi_2 \,, 
\end{align}
where 
\begin{align}
\label{}
\tau\equiv \sqrt{\frac{18\alpha^2}{9\alpha^2+\beta^2}}\kappa \,, \qquad 
\kappa \equiv \sqrt{\kappa_1^2+\kappa_2^2} \,. 
\end{align}
Substitution of this into $S_{11}=-S_{22}$ leads to  
\begin{align}
\label{}
\eta_1=\frac{\beta \tau}{6\alpha}(3 e^{-\phi_1}+e^{\phi_1} )+\frac{3\alpha}{\beta }\tau_3 \,. 
\end{align}
From $\beta (S_{11}+\alpha)-9\alpha^2(S_{12}+\beta)=0$, we get 
\begin{align}
\label{}
(\alpha^2+\beta^2)\tau^2+ 8 \alpha^2 e^{\phi_1}\tau_3\tau=0 \,. 
\end{align}
Supposed $\tau \ne 0$, this equation gives $\tau_3=-e^{-\phi_1}(\alpha^2+\beta^2)\tau /(8\alpha^2)$. 
Then the rest of the Ricci identity  reduces to 
\begin{subequations}
\begin{align}
\label{}
0=&\, 4\alpha\beta[2\beta^2+e^{2\phi_1}(9\alpha^2+\beta^2)]\cos\phi_2- (27\alpha^4+18\alpha^2\beta^2-\beta^4)\sin\phi_2 \,, \\ 
0=&\,  (27\alpha^4+18\alpha^2\beta^2-\beta^4)\cos\phi_2 -4\alpha \beta [-2\beta^2+e^{2\phi_1}(9\alpha^2+\beta^2)]\sin\phi_2\,, \\
0=&\, 3e^{2\phi_1}[\alpha (9\alpha^2+5\beta^2)\cos(2\phi_2)+\beta (\beta^2-3\alpha^2)\sin(2\phi_2)] \notag \\
 &+(9\alpha^2+\beta^2)\left[3\alpha e^{4\phi_1}+\frac{9\alpha^2+\beta^2}{\kappa^2}e^{2\phi_1} -\frac{9(3\alpha^2-\beta^2)(\alpha^2+\beta^2)}{16\alpha \beta^2}\right]\,.
\end{align}
\end{subequations}
One can verify that these equations are not satisfied simultaneously, leading to 
$\tau=\kappa=0$. 
The Jordan form (\ref{zzb1_Segre}) then requires $\alpha =\beta ^2/(6\tau_3^2)$ and 
the nonvanishing Ricci rotation coefficients are given by 
\begin{align}
\label{}
\eta_1=\eta_2 =\frac{\beta}{2\tau_3} \,, \qquad \tau_3 \ne 0 \,. 
\end{align}

The adapted coordinate system is  found by 
inspecting (\ref{eq:app_comm_ortho}), implying the existence of ($t, x,y$) such that 
\begin{align}
\label{}
e_1{}^a=&\, (\partial_t)^a \,, \qquad 
e_2{}^a=(\partial_x)^a \,, \notag \\
e_3{}^a=& \, \left(-\tau_3 x+\eta _1 t+\omega_1(y) \right) (\partial_t)^a
+\left(-\tau_3 t-\eta _1x+\omega_2 (y) \right)(\partial_x)^a
+\omega_3(y)(\partial_y)^a\,, 
\end{align}
where $\omega_{1-3}(y)$ are arbitrary functions of $y$. 
By $\int \D y/\omega_3(y) \to  y $, one can set $\omega_3(y)=1$. 
Functions $\omega_{1,2}(y)$ can be eliminated by the shift 
$t\to t+g_1(y)$, $x\to x+g_2(y)$ where $g_{1,2}$ are chosen to be 
$g_1'+\tau_3 g_2 -\eta _1g_1=\omega_1$ and 
$g_2'+\tau_3 g_1 +\eta _1g_2=\omega_2$. 
It follows that the metric is cast into 
\begin{align}
\label{metric_zz*1}
\D s^2 = -\left[\D t +\left(-\eta_1 t +\tau_3 x\right)\D y\right]^2+
\left[\D x +\left(\eta_1x +\tau_3 t\right)\D y\right]^2 +\D y^2 \,. 
\end{align}
Apart from (\ref{[1,11]metric:(A-ii)constraints}), this metric is obtained by $\tau_1 \to 0$, $\eta_2 \to \eta_1$ of (\ref{[1,11]metric:(A-ii)}). 
The 3 KVs are given by 
\begin{align}
\label{}
K_1 =\frac{1}{\sqrt{\eta_1^2+\tau_3^2}}\partial_y \,, \qquad 
K_\pm = e^{\pm y\sqrt{\eta_1^2+\tau_3^2}} \left[\tau_3 \partial_x - \left(\eta_1 \pm \sqrt{\eta_1^2+\tau_3^2}\right)\partial_t \right]\,, 
\end{align}
with the $\mathfrak e^{1,1}$ algebra ($k=0$ in (\ref{EucFLRW_algebra}))
\begin{align}
\label{}
[ K_1, K_\pm ] =\pm K_\pm \,. 
\end{align}
It has been pointed out recently that this class of stress-energy tensor which is a source of Einstein's equations for the metric (\ref{metric_zz*1})  violates the null energy condition~\cite{Maeda:2018hqu}.
It would be interesting to explore the quantum origin of this  stress-energy tensor.

\appendix 

\section{Curvature and connections}
\label{sec:app}

\subsection{Orthonormal frame}\label{sec:app:ortho}

The metric in the orthonormal frame is given by 
\begin{align}
\label{nonnull_frame}
g_{ab}=\varepsilon e_{1a} e_{1a}+ e_{2a}e_{2b}-\varepsilon e_{3a}e_{3b} \,,
\end{align}
where
\begin{align}
\label{}
\varepsilon =e_1{}^ae_{1a} =-e_3{}^a e_{3a} =\pm 1\,, \qquad 
e_2{}^a e_{2a}=1 \,, \qquad e_1{}^a e_{2a}= e_1{}^a e_{3a}=e_2{}^a e_{3a}= 0\,.
\end{align}
The Ricci rotation coefficients $\{\kappa_i, \eta_i , \tau_i \}$ are defined by 
\begin{subequations}
	\label{eq:inhom1:RRC}
	\begin{align}
	\kappa_1 &~\equiv~ e_1{}^a e_2{}^b \nabla_a e_{1b}\:,
	&\eta_1 &~\equiv~ e_1{}^a e_3{}^b \nabla_a e_{1b}\:,
	&\tau_1 &~\equiv~ e_1{}^a e_3{}^b \nabla_a e_{2b}\:,\\
	%-----
	\kappa_2 &~\equiv~ e_2{}^a e_1{}^b \nabla_a e_{2b}\:,
	&\eta_2 &~\equiv~ e_2{}^a e_3{}^b \nabla_a e_{2b}\:,
	&\tau_2 &~\equiv~ e_2{}^a e_3{}^b \nabla_a e_{1b}\:,\\
	%-----
	\kappa_3 &~\equiv~ e_3{}^a e_1{}^b \nabla_a e_{3b}\:,
	&\eta_3 &~\equiv~ e_3{}^a e_2{}^b \nabla_a e_{3b}\:,
	&\tau_3 &~\equiv~ e_3{}^a e_2{}^b \nabla_a e_{1b}\:,
	\end{align}
\end{subequations}
implying the relation 
\begin{subequations}
\label{RRC_nonull}
\begin{align}
\label{}
\nabla_b e_{1a}=& e_{1b}(\varepsilon \kappa_1 e_{{2}a}-\eta_1 e_{3a})+e_{2b}(-\kappa_2 e_{2a}-\varepsilon \tau_2 e_{3a}) 
%\notag \\&
 +e_{3b}(-\varepsilon \tau_3 e_{2a}-\kappa_3 e_{3a})\,,  \\ 
\nabla_b e_{2a}=& e_{1b}(-\kappa_1 e_{1a}-\tau_1 e_{3a})+\varepsilon e_{2b}(\kappa_2 e_{1a}-\eta_2 e_{3a}) 
%\notag \\&
+e_{3b}(\tau_3 e_{1a}-\eta_3 e_{3a} ) \,, \\
\nabla_{b} e_{3a}=& e_{1b}(-\eta_1 e_{1a}-\varepsilon \tau_1 e_{2a})
+e_{2b}(-\varepsilon \tau_2 e_{1a}-\eta_2 e_{2a})+e_{3b}(-\kappa_3 e_{1a}
-\varepsilon \eta_3 e_{2a} ) \,. 
\end{align}
\end{subequations} 
These basis obeys the following commutation relations
\begin{subequations}
	\begin{align}
	\left[ e_1,e_2 \right]^a &~=~-\varepsilon\:\kappa_1\:e_1^a+\kappa_2\:e_2^a-\varepsilon (\tau_1-\tau_2)e_3^a\:,\\
	%-------
	\left[ e_2,e_3 \right]^a &~=~-\varepsilon(\tau_2-\tau_3) e_1^a-\eta_2\:e_2^a-\varepsilon\:\eta_3\:e_3^a\:,\\
	%-------
	\left[ e_3,e_1 \right]^a &~=~
	\varepsilon\:\eta_1\:e_1^a+(\tau_1+\tau_3)e_2^a+\varepsilon\:\kappa_3\:e_3^a\:.
	\end{align}
	\label{eq:app_comm_ortho}
\end{subequations}
The frame components $R_{ij}=R_{ab}e_i{}^a e_j{}^b$ of the Ricci tensor are given by 
\begin{subequations}
	\begin{align}
	R_{11} ~=~&\eta_1^2 + \varepsilon \eta_1 \eta_2-\kappa_3^2
	+\varepsilon \eta_3 \kappa_1 - \varepsilon \kappa_1^2 -\kappa_2^2
	+2\varepsilon \tau_2 \tau_3
	-\varepsilon \ma L_1 \kappa_3 - \varepsilon\ma L_3 \eta_1 + \ma L_1 \kappa_2 + \ma L_2 \kappa_1\:,\\
	%-----
	R_{22} ~=~& -\eta_3^2 + \eta_1 \eta_2 + \varepsilon \eta_2^2
	-\kappa_1^2 + \kappa_2 \kappa_3 - \varepsilon \kappa_2^2 - 2\tau_3 \tau_1 - \varepsilon \ma L_3 \eta_2
	+\varepsilon \ma L_1 \kappa_2 - \varepsilon \ma L_2 \eta_3 + \varepsilon \ma L_2 \kappa_1\:,\\
	%-----
	R_{33} ~=~&
	\varepsilon \eta_3^2 - \eta_1^2 - \eta_2^2 + \kappa_3^2 - \varepsilon \eta_3 \kappa_1
	-\varepsilon \kappa_3 \kappa_2 - 2\varepsilon \tau_1 \tau_2 + \varepsilon \ma L_3 \eta_1 + \ma L_3 \eta_2
	+\varepsilon \ma L_1 \kappa_3 + \ma L_2 \eta_3 \:,\\
	%-----
	R_{12} ~=~&
	-\eta_3 \kappa_3 - \varepsilon \eta_3 \kappa_2 - \eta_1 \tau_3 + \varepsilon \eta_2 \tau_3 + \eta_1 \tau_2 + \varepsilon \eta_2 \tau_2
	-\varepsilon \ma L_3 \tau_2 - \varepsilon \ma L_2 \kappa_3\:,\\
	~=~&-\eta_3 \kappa_3 - \kappa_3 \kappa_1 - \eta_1 \tau_3 + \varepsilon \eta_2 \tau_3 + \eta_1 \tau_1
	+\varepsilon \eta_2 \tau_1 - \varepsilon \ma L_3 \tau_1 - \varepsilon \ma L_1 \eta_3\:,\\
	%-----
	R_{23} ~=~& -\eta_3 \eta_1 - \eta_1 \kappa_1 -\kappa_3 \tau_3 + \varepsilon \kappa_2 \tau_3 + \kappa_3 \tau_1
	+\varepsilon \kappa_2 \tau_1 + \varepsilon \ma L_3 \kappa_1 - \varepsilon \ma L_1 \tau_3\:,\\
	~=~& -\eta_1 \kappa_1 + \varepsilon \eta_2 \kappa_1 + \kappa_3 \tau_1 + \varepsilon \kappa_2 \tau_1
	-\kappa_3 \tau_2 + \varepsilon \kappa_2 \tau_2 - \varepsilon \ma L_1 \tau_2 + \varepsilon \ma L_2 \eta_1\:,\\
	%-----
	R_{31} ~=~& -\varepsilon \eta_2 \kappa_3 - \eta_2 \kappa_2 + \varepsilon \eta_3 \tau_3 - \varepsilon \kappa_1 \tau_3
	+\varepsilon \eta_3 \tau_2 + \varepsilon \kappa_1 \tau_2 + \ma L_3 \kappa_2 + \ma L_2 \tau_3\:,\\
	~=~& \varepsilon \eta_1 \kappa_2 - \eta_2 \kappa_2 - \varepsilon \eta_3 \tau_1 + \varepsilon \kappa_1 \tau_1
	+\varepsilon \eta_3 \tau_2 + \varepsilon \kappa_1 \tau_2 + \ma L_1 \eta_2 -\ma L_2 \tau_1\:.
	\end{align}
	\label{eq:app_ricci_ortho}
\end{subequations}
Here and throughout the body of text, we denote 
$\ma L_i =\ma L_{e_i}$ for the Lie derivative along $e_i$.

There are three kinds of transformations for the basis corresponding to local 
${\rm SO}(2,1)$: 
\begin{subequations}
\label{nonnull_localSO21}
\begin{align}
\label{}
({\rm I}) & ~ e_{1a} \to \cos (\sqrt\varepsilon a_1) e_{1a}+\frac 1{\sqrt\varepsilon}
\sin (\sqrt\varepsilon a_1) e_{2a} \,, \qquad 
e_{2a} \to -\sqrt\varepsilon \sin (\sqrt\varepsilon a_2) e_{1a} 
+\cos (\sqrt\varepsilon a_1)e_{2a} \,, \qquad 
e_{3a} \to e_{3a} \,,\label{nonnull_localSO21_I} \\
({\rm II}) & ~e_{1a} \to e_{1a} \,, \quad ~
e_{2a} \to \cosh (\sqrt\varepsilon a_2) e_{2a} -\sqrt\varepsilon \sinh (\sqrt\varepsilon a_2)e_{3a} \,, \quad ~
e_{3a} \to -\frac {\sinh (\sqrt\varepsilon a_2)}{\sqrt\varepsilon}e_{2a}+\cosh (\sqrt\varepsilon a_2)
e_{3a} \,, \label{nonnull_localSO21_II}\\ 
({\rm III}) & ~ e_{1a}  \to \cosh (a_3)e_{1a}+\sinh (a_3)e_{3a} \,, \qquad 
e_{2a} \to e_{2a} \,, \qquad 
e_{3a} \to \sinh (a_3) e_{1a}+ \cosh (a_3) e_{3a} \,,\label{nonnull_localSO21_III}
\end{align}
\end{subequations} 
where $a_{1,2,3}$ are arbitrary functions. 
The Ricci rotation coefficients transform respectively as 
\begin{subequations}
\label{nonnull_RRC_localSO21_I}
\begin{align}
\label{}
\kappa_1 &\to \cos(\sqrt \varepsilon a_1) (\kappa_1 +\ma L_1 a_1)
+\frac 1{\sqrt \varepsilon } \sin (\sqrt \varepsilon  a_1)(-\kappa_2+\ma L_2 a_1)\,,  \\ 
\eta_1 & \to \cos^2 (\sqrt \varepsilon a_1) \eta_1+\varepsilon  \sin^2(\sqrt \varepsilon  a_1) \eta_2+\frac{1}{\sqrt \varepsilon }\sin(\sqrt \varepsilon a_1)\cos(\sqrt \varepsilon a_1) (\tau_1+\tau_2)\,, \\
\tau_1 &\to \cos ^2 (\sqrt \varepsilon a_1)\tau_1-\sin^2(\sqrt \varepsilon a_1)\tau_2
+\frac 1{\sqrt \varepsilon } \sin(\sqrt \varepsilon a_1)\cos(\sqrt \varepsilon a_1)(\eta_2- \varepsilon  \eta_1)\,, \\
\kappa_2 &\to \cos (\sqrt \varepsilon a_1)(\kappa_2-\ma L_2 a_1)+\sqrt \varepsilon 
\sin(\sqrt \varepsilon a_1)(\kappa_1+\ma L_1 a_1) \,,  \\
\eta_2& \to \cos^2 (\sqrt \varepsilon a_1) \eta_2 + \varepsilon \sin^2(\sqrt \varepsilon a_1)
\eta_1 -\sqrt \varepsilon \sin(\sqrt \varepsilon a_1)\cos(\sqrt \varepsilon a_1)(\tau_1+\tau_2)\,, \\ 
\tau_2 &\to \cos^2(\sqrt \varepsilon a_1)\tau_2-\sin^2(\sqrt \varepsilon a_1)\tau_1
+\frac 1{\sqrt \varepsilon }\sin(\sqrt \varepsilon a_1)\cos(\sqrt \varepsilon a_1)(\eta_2- \varepsilon  \eta_1) \,,  \\
\kappa_3 &\to  \cos(\sqrt \varepsilon a_1)\kappa_3+\frac 1{\sqrt \varepsilon }\sin(\sqrt \varepsilon a_1)\eta_3 \,, \\ 
\eta_3 &\to \cos (\sqrt \varepsilon a_1)\eta_3 -\sqrt \varepsilon \sin(\sqrt \varepsilon a_1) \kappa_3 \,, \\
\tau_3 &\to \tau_3+\ma L_3 a_1 \,, 
\end{align}
\end{subequations} 

\begin{subequations}
\label{nonnull_RRC_localSO21_II}
\begin{align}
\label{}
\kappa_1 &\to \cosh(\sqrt \varepsilon a_2) \kappa_1 -\sqrt \varepsilon \sinh(\sqrt \varepsilon a_2)\eta_1 \,, \\
\eta_1 & \to \cosh (\sqrt \varepsilon a_2) \eta_1-\frac 1{\sqrt \varepsilon}  \sinh (\sqrt \varepsilon  a_2) \kappa_1 \,, \\
\tau_1 &\to \tau_1+\ma L_1 a_2\,,   \\
\kappa_2 &\to \cosh^2 (\sqrt \varepsilon a_2)\kappa_2+\varepsilon \sinh^2(\sqrt \varepsilon a_2) \kappa_3 
+\sqrt \varepsilon  \sinh(\sqrt \varepsilon a_2)\cosh(\sqrt \varepsilon a_2)(\tau_2+\tau_3) \,,  \\
\eta_2& \to
\cosh(\sqrt \varepsilon a_2)(\eta_2+\ma L_2 a_2)+\sqrt \varepsilon\sinh(\sqrt \varepsilon a_2)(\eta_3-\ma L_3 a_2) \,,  \\
\tau_2 &\to \cosh^2(\sqrt \varepsilon a_2)\tau_2+\sinh^2(\sqrt \varepsilon a_2)\tau_3
+\frac 1{\sqrt \varepsilon}\sinh(\sqrt \varepsilon a_2)\cosh (\sqrt \varepsilon a_2)(\kappa_2+\varepsilon \kappa_3)\,,  \\
\kappa_3 &\to  \cosh^2 (\sqrt \varepsilon a_2)\kappa_3+ \varepsilon \sinh^2 (\sqrt \varepsilon a_2)\kappa_2+\frac 1{\sqrt \varepsilon}
\sinh(\sqrt \varepsilon a_2)\cosh(\sqrt \varepsilon a_2)(\tau_2+\tau_3) \,, \\ 
\eta_3 &\to \cosh(\sqrt \varepsilon a_2)(\eta_3-\ma L_3 a_2)+\frac 1{\sqrt \varepsilon}\sinh(\sqrt \varepsilon a_2) (\eta_2+\ma L_2 a_2 ) \,, \\
\tau_3 &\to \cosh^2 (\sqrt \varepsilon a_2)\tau_3+\sinh^2(\sqrt \varepsilon a_2)\tau_2 +\frac 1{\sqrt \varepsilon}
\sinh(\sqrt \varepsilon a_2)\cosh (\sqrt \varepsilon a_2)(\kappa_2+\varepsilon \kappa_3) \,, 
\end{align}
\end{subequations}

\begin{subequations}
\label{nonnull_RRC_localSO21_III}
\begin{align}
\label{}
\kappa_1 &\to \cosh^2 (a_3) \kappa_1+ \sinh^2 (a_3) \eta_3+\cosh (a_3) \sinh (a_3) (\tau_3-\tau_1) \,, \\
\eta_1 & \to\cosh(a_3)(\eta_1-\varepsilon \ma L_1 a_3)-\sinh(a_3)(\kappa_3 +\varepsilon \ma L_3 a_3 )\,, \\
\tau_1 &\to \cosh^2(a_3 )\tau_1 -\sinh^2 (a_3 ) \tau_3 -\sinh(a_3)\cosh(a_3)(\kappa_1+\eta_3)\,, \\
\kappa_2 &\to  \cosh(a_3)\kappa_2 +\sinh(a_3 )\eta_2\,,  \\
\eta_2& \to \cosh(a_3)\eta_2 +\sinh(a_3 ) \kappa_2 \,,  \\
\tau_2 &\to \tau_2- \varepsilon \ma L_2 a_3 \,,  \\
\kappa_3 &\to  \cosh(a_3)(\kappa_3 +\varepsilon \ma L_3 a_3) +\sinh(a_3) (-\eta_1+\varepsilon \ma L_1 a_3 )\,, \\ 
\eta_3 &\to \cosh^2(a_3) \eta_3 +\sinh^2(a_3 ) \kappa_1 +\sinh(a_3)\cosh(a_3 )(\tau_3-\tau_1) \,, \\
\tau_3 &\to \cosh^2 (a_3 )\tau_3 -\sinh^2(a_3) \tau_1+\sinh(a_3)\cosh(a_3)(\kappa_1+\eta_3) \,. 
\end{align}
\end{subequations}

\subsection{Null frame}

In the null frame, the metric is given by 
\begin{align}
\label{nullframe}
g_{ab}=2u_{(a}v_{b)} +e_a e_b \,, 
\end{align}
where 
\begin{align}
\label{}
g_{ab} u^au^b = g_{ab}v^a v^b=g_{ab}u^a e^b=g_{ab}v^a e^b=0 \,, \qquad 
g_{ab}u^a v^b=g_{ab}e^a e^b=1 \,. 
\end{align}
The 9 Ricci rotation coefficients are defined by 
\begin{subequations}
\label{null:RRC}
	\begin{align}
	\kappa_u &~\equiv~ v^a u^b \nabla_b u_a \:,
	&\eta_u &~\equiv~ e^a u^b \nabla_b u_{a}\:,
	&\tau_u &~\equiv~ e^a u^b \nabla_b v_a\:,\\
	%-----
	\kappa_v &~\equiv~ u^a u^b \nabla_b v_a \:,
	&\eta_v &~\equiv~ e^a v^b \nabla_b v_a\:,
	&\tau_v &~\equiv~ e^a v^b \nabla_b u_a \:,\\
	%-----
	\kappa_e &~\equiv~ u^a e^b \nabla_b e_{a}\:,
	&\eta_e &~\equiv~ v^a e^b \nabla_a e_{b}\:,
	&\tau_e &~\equiv~  v^a e^b \nabla_b u_a\:,
	\end{align}
\end{subequations}
leading to
\begin{subequations}
\begin{align}
\label{}
\nabla_b u_a =& \kappa_u u_a v_b+\eta_u e_a v_b-\kappa_v u_a u_b 
+\tau_ve_a u_b-\kappa_e e_a e_b+\tau_e u_a e_b \,, \\
\nabla_b v_a =& -\kappa_u v_a v_b +\tau_u e_a v_b +\kappa_v v_a u_b 
+\eta_v e_a u_b -\eta_e e_a e_b -\tau_e v_a e_b \,, \\
\nabla_b e_a =& -\eta_u v_a v_b -\tau_u u_a v_b -\eta_vu_a u_b 
-\tau_v v_a u_b+\kappa_e v_a e_b +\eta_e u_a e_b \,. 
\end{align}
\end{subequations}
The commutation relations are 
\begin{subequations}
	\begin{align}
	\left[ u,v \right]^a &~=~
	\kappa_v\:u^a -\kappa_u\:v^a + (\tau_u-\tau_v) e^a
	\:,\\
	%-------
	\left[ v,e \right]^a &~=~
	-\eta_v\:u^a - (\tau_v - \tau_e) v^a + \eta_e\:e^a\:,\\
	%-------
	\left[ e,u \right]^a &~=~
	(\tau_u+\tau_e)u^a + \eta_u\:v^a -\kappa_e\:e^a\:.
	\end{align}
	\label{eq:app_comm_null}
\end{subequations}
The Ricci tensor is projected to 
\begin{subequations}
	\begin{align}
	R_{uu} ~=~&
	-\kappa_e^2-\kappa_e\kappa_u -2\eta_u \tau_e -\eta_u \tau_u -\eta_u \tau_v
	+\ma L_e \eta_u + \ma L_u\kappa_e\:,\\
	%------
	R_{vv}~=~&
	-\eta_e^2 - \eta_e \kappa_v + 2\eta_v \tau_e - \eta_v \tau_u - \eta_v \tau_v + \ma L_e \eta_v +\ma L_v \eta_e\:,\\
	%------
	R_{ee} ~=~&
	-2\eta_u \eta_v -2 \eta_e \kappa_e + \eta_e \kappa_u + \kappa_e \kappa_v
	-\tau_u^2 -\tau_v^2 + \ma L_e \tau_u + \ma L_e \tau_v + \ma L_u \eta_e + \ma L_v \kappa_e\:,\\
	%------
	R_{uv} ~=~&
	-\eta_e \kappa_e + \kappa_e \kappa_v - 2 \kappa_u \kappa_v - \tau_e \tau_u + \tau_e \tau_v - \tau_u \tau_v -\tau_v^2
	+\ma L_e \tau_v - \ma L_u \kappa_v + \ma L_v \kappa_e - \ma L_v \kappa_u\:,\\
	~=~&
	-\eta_e \kappa_e + \eta_e \kappa_u - 2 \kappa_u \kappa_v - \tau_e \tau_u - \tau_u^2 + \tau_e \tau_v - \tau_u \tau_v
	+\ma L_e \tau_u + \ma L_u \eta_e - \ma L_u \kappa_v - \ma L_v \kappa_u\:,\\
	%------
	R_{ve} ~=~&
	-\eta_v \kappa_e - \eta_v \kappa_u + \eta_e \tau_e - \kappa_v \tau_e + \eta_e \tau_v + \kappa_v \tau_v -\ma L_e \kappa_v -\ma L_v \tau_e\:,\\
	%------
	~=~& -2\eta_v \kappa_u -\eta_e \tau_u + \eta_e \tau_v - \ma L_u \eta_v + \ma L_v \tau_u\:,\\
	%------
	R_{ue} ~=~&
	-\eta_e \eta_u - \eta_u \kappa_v - \kappa_e \tau_e + \kappa_u \tau_e + \kappa_e \tau_u + \kappa_u \tau_u
	-\ma L_e \kappa_u + \ma L_u \tau_e\:,\\
	~=~& -2\eta_u \kappa_v + \kappa_e \tau_u - \kappa_e \tau_v + \ma L_u \tau_v - \ma L_v \eta_u\:.
	\end{align}
	\label{eq:app_ricci_null}
\end{subequations}
where 
\begin{align}
\label{}
R_{uu}=&\, R_{ab} u^a u^b\,, \qquad 
R_{uv}=R_{ab} u^a v^b\,, \qquad 
R_{ue}=R_{ab} u^a e^b\,, \notag \\
R_{vv}=&\, R_{ab} v^a v^b\,, \qquad 
R_{ve}=R_{ab} v^a e^b\,, \qquad 
R_{ee}=R_{ab} e^a e^b\,. 
\end{align}
Throughout the paper, we adopt the same notation for the trace-free Ricci tensor, 
e.g, $S_{uu}=S_{ab}u^a u^b$.

The 3-kinds of local Lorentz transformations are expressed as 
\begin{subequations}
\label{null_localSO21}
\begin{align}
\label{null_localSO21_I}
({\rm I}) ~~ &u_a \to b_1 u_a \,, \qquad v_a \to b_1^{-1}v_a \,, \qquad e_a \to e_a \,, \\
({\rm II}) ~~ &u_a \to u_a \,, \qquad  v_a \to v_a -\frac 12 b_2^2 u_a +b_2 e_a \,,\label{null_localSO21_II} \qquad 
e_a \to e_a -b_2 u_a  \,, \\
\label{null_localSO21_III}
({\rm III}) ~~ &u_a \to  u_a -\frac 12 b_3^2 v_a +b_3 e_a \,, \qquad 
v_a \to v_a \,, \qquad 
e_a \to e_a -b_3 v_a \,,
\end{align}
\end{subequations}
where $b_1,b_2, b_3$ are arbitrary functions. 
Under these frame change, the transformation rules for the Ricci rotation coefficients are summarized as
\begin{align}
\label{}
\kappa_u \to b_1 \kappa_u +\ma L_u b_1 \,, \qquad \eta_u \to b_1^2 \eta_u \,, \qquad \tau_u\to \tau_u \qquad 
\kappa_v \to b_1^{-1}\kappa_v -b_1^{-2}\ma L_v b_1 \,, \notag \\
\tau_v \to \tau_v \,, \qquad \eta_v \to b_1^{-2} \eta_ v \, \qquad \tau_e \to \tau_e +\ma L_e \log b_1 \,, \qquad 
\kappa_e \to b_1 \kappa_e \,, \qquad \eta_e \to b_1^{-1} \eta_e \,.
\end{align}

\begin{align}
\label{}
\kappa_u \to & \kappa_u +b_2 \eta_u \,, \qquad \eta_u \to \eta_u \,, \qquad \tau_u \to \tau_u +b_2 \kappa_u +\frac 12 b_2^2 \eta_u +\ma L_u b_2 \,, \notag \\
\kappa_v \to & \kappa_v-b_2(\tau_e+\tau_v)+b_2^2 \left(\kappa_e+\frac 12 \kappa_u \right)+\frac 12 b_2^3 \eta_u \,, \qquad 
\tau_v \to \tau_v -b_2 \kappa_e -\frac 12 b_2^2 \eta_u \,, \notag \\ 
\eta_v \to & \eta_v+\ma L_v b_2 +b_2(-\eta_e -\kappa_v+\ma L_e b_2 ) +\frac 12 b_2^2(2\tau_e-\tau_u +\tau_v -\ma L_u b_2)
-\frac 12 b_2^3(\kappa_e +\kappa_u)-\frac 14 b_2^4 \eta_u \,, \notag \\ 
\tau_e \to & \tau_e -b_2(\kappa_e+\kappa_u)-b_2^2 \eta_u \,, \qquad 
\kappa_e \to \kappa_e +b_2 \eta_u \,, \notag \\
\eta_e \to & \eta_e +b_2 (-\tau_e+\tau_u+\ma L_u b_2)
-\ma L_e b_2 +b_2^2 \left(\frac 12 \kappa_e +\kappa_u \right)+\frac 12 b_2^3 \eta_u \,.
\end{align}

\begin{align}
\label{}
\kappa_u \to &  \kappa_u +b_3(\tau_u-\tau_e)+b_3^2 \left(\eta_e+\frac 12 \kappa_v \right) -\frac 12 b_3^3 \eta_v \,, \qquad 
\tau_u \to \tau_u +b_3 \eta_e -\frac 12 b_3^2 \eta_v \,, \notag \\
\eta_u \to & \eta_u +b_3(\kappa_e+\kappa_u +\ma L_e b_3)-\ma L_u b_3 +\frac 12 b_3^2 (\tau_u-\tau_v -2\tau_e+\ma L_v b_3 ) 
+\frac 12 b_3^3 (\eta_e +\kappa_v)-\frac 14 b_3^4 \eta_v \,, \notag \\ 
\kappa_v \to & \kappa_v -b_3 \eta_v  \,, \qquad \eta_v \to \eta_v \,, \qquad 
\tau_e \to \tau_e -b_3(\eta_e+\kappa_v)+b_3^2 \eta_v \,, \notag \\
\kappa_e \to & \kappa_e +b_3(-\tau_e-\tau_v +\ma L_v b_3 ) +\ma L_e b_3 +b_3^2 \left(\frac 12 \eta_e +\kappa_v \right)
-\frac 12 b_3^3 \eta_v \,, \qquad \eta_e \to \eta_e -b_3 \eta_v \,.  
\end{align}

\acknowledgments
This work is partially supported by 
Grant-in-Aid for Scientific Research (A) from JSPS 17H01091.

%%%%%%%%%%%%%%%%%%%%%%%%%%%%%%%
%                                                                                                %
%                                   References                                         %
%                                                                                                %
%%%%%%%%%%%%%%%%%%%%%%%%%%%%%%%

\end{document}